\documentclass[fleqn,usenatbib]{mnras}

\usepackage[T1]{fontenc}
\usepackage{ae,aecompl}

\usepackage{graphicx}	
\usepackage{amsmath}	
\usepackage{amssymb}	
\usepackage{indentfirst}
\usepackage{longtable}  
\usepackage{placeins}   
\usepackage{color}     
\usepackage[symbol]{footmisc}
\usepackage{fp}
\usepackage{multirow}
\usepackage{footnote}
\usepackage{natbib}
\usepackage{hyperref}
\hypersetup{
    colorlinks=true,
    linkcolor=blue,
    filecolor=magenta,
    urlcolor=blue,
}

\newcommand{\lt}{<}
\newcommand{\gt}{>}

\title[Tidal star-planet interaction]{Tidal star-planet interaction and its observed impact on stellar activity in planet-hosting wide binary systems}

\author[N. Ilic et al.]{
N. Ilic$^{1, 2}$\thanks{E-mail: nilic@aip.de},
K. Poppenhaeger$^{1, 2}$,
and S. Marzieh Hosseini$^{1, 2}$
\\
$^{1}$Leibniz Institute for Astrophysics Potsdam (AIP), An der Sternwarte 16, 14482 Potsdam, Germany\\
$^{2}$Universit\"at Potsdam,  Institut f\"ur Physik und Astronomie,  Karl-Liebknecht-Stra\ss e 24/25, 14476 Potsdam, Germany}

\date{Accepted xxx. Received xxx; in original form xxx.}

\pubyear{2021}

\begin{document}
\label{firstpage}
\pagerange{\pageref{firstpage}--\pageref{lastpage}}
\maketitle

\begin{abstract}

Tidal interaction between an exoplanet and its host star is a possible pathway to transfer angular momentum between the planetary orbit and the stellar spin. In cases where the planetary orbital period is shorter than the stellar rotation period, this may lead to angular momentum being transferred into the star's rotation, possibly counteracting the intrinsic stellar spin-down induced by magnetic braking. Observationally, detecting altered rotational states of single, cool field stars is challenging, as precise ages for such stars are rarely available. Here we present an empirical investigation of the rotation and magnetic activity of a sample of planet-hosting stars that are accompanied by wide stellar companions. Without needing knowledge about the absolute ages of the stars, we test for relative differences in activity and rotation of the planet hosts and their co-eval companions, using X-ray observations to measure the stellar activity levels. Employing three different tidal interaction models, we find that host stars with planets that are expected to tidally interact display elevated activity levels compared to their companion stars. We also find that those activity levels agree with the observed rotational periods for the host stars along the usual rotation-activity relationships, implying that the effect is indeed caused by a tidal interaction and not a purely magnetic interaction which would be expected to affect the stellar activity, but not necessarily the rotation. We conclude that massive, close-in planets have an impact on the stellar rotational evolution, while the smaller, more distant planets do not have a significant influence.

\end{abstract}




\section{Introduction}

Since the discovery of the first extra-solar planet \citep{MayorQueloz1995}, the interaction between a star and its planetary-mass companion, and their common evolution, have been a topic of interest \citep{Rasio1996,Cuntz2000,Bodenheimer2001,Oglivie2004,Dobbs-Dixon2004,Shkolnik2005}. The idea of a star-planet interaction comes from the interaction between stellar components in close binary systems, i.e.\ one regards the star-planet system as a binary system with a very uneven mass ratio \citep{Zahn1977,Hut1981}. The main pathways for star-planet interaction (SPI) are magnetic and tidal interactions. Which of the two types of interaction dominates depends mostly on the initial configuration of the system and the evolutionary phase considered \citep{Ahuir2021}.

One of the first explorations of the tidal and magnetic interaction was the work by \cite{Cuntz2000}, who considered tidal interaction in the form of tidal bulges raised by the planet on the stellar photosphere, and magnetic interaction in the form of interaction between the planetary magnetosphere and the magnetic field of active regions. In the following works, magnetic and tidal interaction types have been explored in more detail.

In most cases, the effects of magnetic star-planet interaction are thought to happen on short time scales. This type of interaction can produce flaring and other types of high-energy events in the process of magnetic field line reconnection \citep{Ip2004}. One of the first observational investigations of possible increased activity level due to SPI was by \cite{Shkolnik2005}, where the chromospheric activity of 13 planet-hosting stars was monitored. It was found that two of the stars exhibit activity enhancements in phase with the planetary orbital rate, suggesting magnetic instead of tidal interaction. A follow-up study \citep{Shkolnik2008} found variability with the planetary orbit only for one of the two targets, which was interpreted as on/off behavior of SPI. Further investigations targeted single-epoch coronal observations and investigated stellar activity as a function of planetary mass and orbital period \citep{Kashyap2008, Poppenhaeger2010}. The picture is complicated by selection biases that exist for exoplanet detection and are a function of stellar activity themselves \citep{Poppenhaeger2011scharf}. A large sample of chromospheric measurements showed potential star-planet interaction effects on the stellar activity only for the closest and heaviest exoplanet companions \citep{Miller2015}.   


The main aspect of tidal star-planet interaction is thought to be the long-term planetary orbital and stellar spin evolution, where the angular momentum transfer between the two components plays a significant role. 
The orbital evolution is manifested as the change of orbital inclination \citep{Winn2010,Albrecht2012}, and the simultaneous change of the semi-major axis and orbital eccentricity \citep{Jackson2008, Strugarek2017}. One other possibility involves a continuous orbital decay, which may result in the destruction of the planet within the lifetime of the star \citep{Patzold2002,Bolmont2012,Benbakoura2019}.

In general, the spin evolution of a star is determined by its initial spin, contraction rate, and the efficiency of the stellar wind. In particular, the stellar wind is important since it carries away angular momentum from the star. 
This process, called magnetic braking, slows down the rotation rate of the star over timescales of gigayears \citep{Belcher1976, Mestel1968, Kraft1967,WeberDavis1967} and can weaken magnetic phenomena such as starspots, stellar winds, and coronal emission. 
For a star with a close-in massive planet, this scenario can be modified: it is expected that planets with an orbital angular velocity higher than the stellar rotational angular velocity will decrease the spin-down rate of the host star, while their orbit decays, altering the expected stellar rotational evolution (e.g. \cite{Brown2011,Ferraz-Mello2015,Penev2016,Gallet2018}). One of the consequences will be the altered activity level of the star. 

To trace tidal star-planet interaction, different indicators have been explored. The distribution of stellar rotation rates with planetary mass and distance were examined, where the indication for tidal evolution in transiting planetary systems was found \citep{Pont2009}. By calculating the gyrochronological ages of stars using their rotational periods and comparing them to estimated isochrone ages, it was found that the gyro-ages of some transiting exoplanet host stars are significantly lower than the isochrone age estimates, indicating SPI \citep{Maxted2015}.
Another study investigated the galactic velocity dispersion of main-sequence planet-host stars \citep{Hamer2019}. It was shown that the hosts are preferentially younger than a matched sample of field stars which indicated that close-in giant planets are destroyed by tides while their host stars are on the main sequence.

One other commonly used tidal star-planet interaction indicator is the level of stellar activity, in particular, the X-ray luminosity of the stellar corona. Since the X-ray luminosity is a function of stellar rotational rate \citep{Pallavicini1981,Pizzolato2003,Wright2011}, if tidal SPI can alter the rotation rate of the host star and possibly induce a spin-up, a higher X-ray luminosity will be observed.



As discussed by \cite{Poppenhaeger2010} and \cite{Poppenhaeger2011scharf}, using stellar activity as an indicator for tidal star-planet interaction can introduce selection effects. Since stellar activity masks the planet-induced radial-velocity signal, small far-out planets are more easily detected around very inactive stars. Therefore, using the RV method, more massive, close-in planets are the only planet type that is easy to detect around active stars. Similarly, smaller planets are easier to detect around magnetically quiet stars and may be missed in the case of magnetically active stars. This is somewhat ameliorated by transit observations, which are less vulnerable to high stellar activity, and small planets have been found around young and active stars \citep{Newton2019,Plavchan2020}. However, confirming those planet candidates through mass determinations which mainly involve RV measurements stays observationally costly.
Therefore, a given sample of successfully detected exoplanets can be expected to have a detection bias causing an excess of Hot Jupiters known around high-activity stars, which needs to be disentangled from genuine star-planet interaction signatures.

In this study, we utilize the X-ray luminosity of planet-hosting stars as an indicator for stellar activity and avoid the selection effect described above by a particular sample definition. Here, we investigate planet hosts that reside in wide binary systems. Our work, however, differs from that by \cite{Poppenhaeger2014} since we do not compare the estimated age of the binary companions but their measured activity level. Our analysis makes use of the assumption that both components formed at similar times and therefore have similar age, which will leave the activity level of the companions to depend only on their spectral type and possibly on the tidal star-planet interaction strength. After correcting for the spectral class difference, the activity level of the companion star will be used to estimate the expected activity level of the host itself. The secondary does not have a detected planet or has a distant less massive planet. With this approach, we have eliminated the selection effect due to the radial velocity detection method since we do not compare the activity level of the host stars to one another, but an independent activity proxy.

In section \ref{obs_data_analysis}, we described the observational data obtained for our study and the methods used to analyse the data and calculate the X-ray flux and luminosity of our sources. In section \ref{results} the normalised difference in the X-ray luminosity of the binary system components is estimated, and the parameters accounting for the star-planet tidal interaction strength are calculated. Section \ref{discussion} discusses the relation between the X-ray luminosity difference and the tidal interaction strength, accounting for the possible activity biases in our stellar sample and the influence of magnetic star-planet interaction. Section \ref{conclusion} summarises the finding of our study.




\section{Observations and data analysis}
\label{obs_data_analysis}

We constructed a sample of nearby exoplanet host stars in wide common proper motion systems with one or more other stars. Such systems were presented by \cite{Raghavan2010}, who found that $\approx 45\%$ of main-sequence stars of spectral type F6 to K3 have one or more stellar companions and that these statistics also apply to planet-hosting stars.
In a recent work by \cite{Mugrauer2019}, the multiplicity of known planet-hosting stars, in particular, was examined. Using the second data release of the ESA-Gaia mission, the authors found equidistant stellar companions that share a common proper motion with the planet-hosting stars, thereby establishing gravitationally bound systems. In addition to systems from this survey, we have also included the binary system HAT-P-20 \citep{Knutson2014} into our analysis, where the primary component hosts a close-in giant planet.

We acquired our sample by requiring that the two stellar companions have a separation larger than 100 AU\footnote{This value is the lower limit of the range given in the cited research paper. By adopting the upper limit of 300 AU, our final results do not change significantly.} in order to avoid significant star-star interactions during the formation of the system \citep{Desidera2007}. One system which was in our initial sample but was later on excluded is the triple-star system HD~26965 \citep{Ma2018} since the reality of the planet is debated due to the similarity of the stellar rotation period and the potential planet-induced radial velocity signature \citep{Diaz2018}.

We then collected X-ray data of the systems to measure the stellar activity levels by inspecting data of the XMM-Newton and Chandra X-ray observatories from several observing programs\footnote{I.e., XMM-Newton programs 0722030 and 0728970 as well as Chandra program 15200512, PI Poppenhaeger}, as well as archival observations of other systems. This yielded a sample of 34 systems with existing X-ray coverage in XMM-Newton or Chandra.


\subsection{Analysis of XMM-Newton data}
\subsubsection{XMM-Newton instrumentation}

All sources found in the XMM-Newton archive were observed with the telescope's European Photon Imaging Camera (EPIC). EPIC is composed of three cameras, two are Metal Oxide Semiconductor (MOS) CCD arrays and one is a positive-negative junction (PN) CCD array. These two types of cameras differ in their instrument design, the geometry of the CCD arrays, and their readout time. Also, the pn CCDs are back-illuminated, while MOS CCDs are front-illuminated, which affects the chip's quantum efficiency, making the pn camera more sensitive to X-ray photons.

The EPIC cameras perform high sensitivity imaging observations over the telescope's circular field of view (FoV) of $\approx 30\arcmin$ in radius. They are sensitive to the energy range of 0.15-15 keV with a maximum sensitivity at 1.5 keV. The in-flight on-axis point spread function (PSF) of the cameras, at a photon energy of 1.5 keV, yields a full-width-half-maximum (FWHM) of $\approx 6 \arcsec$ for all three cameras \citep{EPICmos,EPICpn}.

For our sample observations, all EPIC CCDs operated in photon counting mode with a fixed frame read-out frequency, producing event lists. Each detected event is associated with the position on the chip at which they were registered, their arrival time, and energies.

Onboard XMM-Newton is also the EPIC Radiation Monitor (ERM), which is employed for monitoring the space radiative and particle environment, which mostly depends on Earth's radiative belts and solar flares, and an Optical Monitor (OM) and Reflection Grating Spectrometers (RGS), but they were not used for our analysis.

In Table \ref{table:1} are listed all systems acquired in the XMM-Newton archive.

\subsubsection{Photon event extraction}

\begin{figure}
  \includegraphics[width=\linewidth, height=8.75cm]{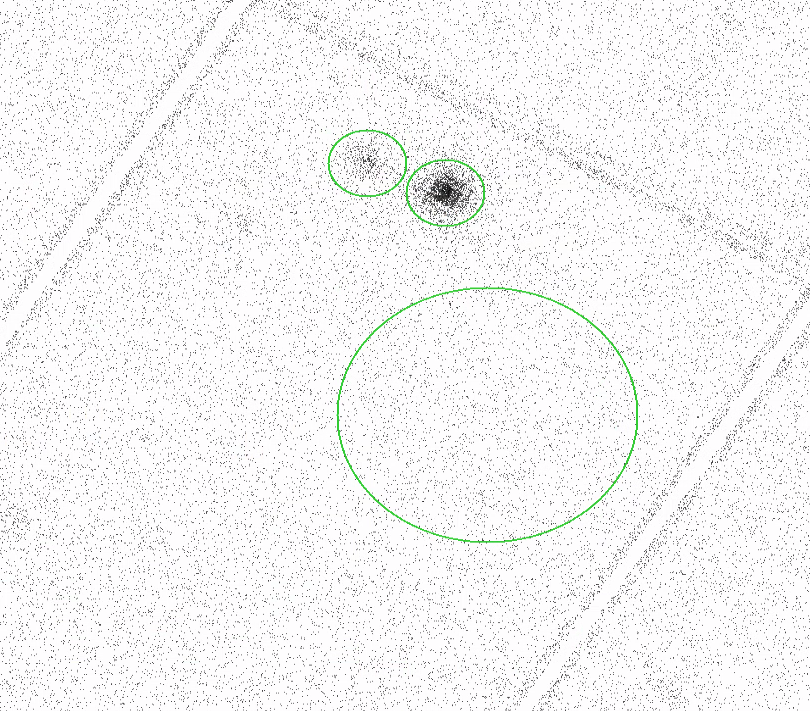}
  \caption{Shown is an example of the source and background extraction regions for the binary system Gliese 15 (GJ-15) on the observation taken with the EPIC pn camera. At the top are two source extraction regions: the region on the right encompasses the planet host, while the region on the left encircles the planet-host companion. The large circle at the center represents the background extraction region. The radius of the two source regions and the background region is $\approx$ 15.8\arcsec and 60\arcsec, respectively.}
  \label{fig:extraction_region}
\end{figure}

To analyze the data collected with EPIC, we first processed the observations with the XMM-Newton Science Analysis System (SAS), version 18.0.0. For each observation, we used the appropriate Current Calibration Files (CCF) and run the EPIC reduction meta-task {\it emproc} and {\it epproc} to reduce the data collected by the MOS and PN camera, respectively. Then, for every reduced observation, we defined a circular source extraction region for the planet host and its companion(s), and a circular, source-free, background extraction region\footnote{The source-free background extraction region was determined by visual inspection. Each observation was inspected and a region where no obvious source was seen has been chosen for estimation of the X-ray background signal.} (see Fig. \ref{fig:extraction_region}). To include most of the camera's PSF, we chose the on-sky source extraction region radius to be between 10$\arcsec$ and 20$\arcsec$, depending on the angular separation between the stellar components. The background extraction region has a radius of 60$\arcsec$, providing a good estimate of the background signal.

\begin{table}
\centering
\begin{tabular}{lccc}
\hline
\hline
system & obs ID & component & camera \\

\hline

\multirow{2}{*}{16 Cyg} & \multirow{2}{*}{0823050101} & AC & all\\
&&B*&all\\[0.2cm]

\multirow{2}{*}{30 Ari} & \multirow{2}{*}{0075940101} & A & all\\
&&B*C&all\\[0.2cm]

\multirow{2}{*}{55 Cnc} & \multirow{2}{*}{0551020801} & A* & all \\
& & B & all\\[0.2cm]

\multirow{2}{*}{83 Leo} & \multirow{2}{*}{0551021201} & A & all\\
& & B* & all\\[0.2cm]

AS 205 & 0602730101 & A*B & all\\[0.2cm]

\multirow{2}{*}{GJ 15} & \multirow{2}{*}{0801400301} & A* & all\\
& & B & all\\[0.2cm]

\multirow{4}{*}{HAT-P-16} & \multirow{2}{*}{0800733101} & A*B & all\\
&&C&all\\
& \multirow{2}{*}{0800730701} & A*B & all \\
&&C&all\\[0.2cm]

\multirow{4}{*}{HD 27442}&\multirow{2}{*}{0780510501}& A* & all\\
&&B&all\\
&\multirow{2}{*}{0551021401}& A* & all\\
&&B&all\\[0.2cm]

HD 46375 & 0304202501 & A*B & all\\[0.2cm]

\multirow{4}{*}{HD 75289} & \multirow{2}{*}{0304200501} & A* & all\\
&&B&all\\
&  \multirow{2}{*}{0722030301} & A* & all\\
&&B&all\\[0.2cm]

\multirow{2}{*}{HD 101930} & \multirow{2}{*}{0555690301} & A* & all\\
& & B & pn \\[0.2cm]

\multirow{2}{*}{HD 107148} & \multirow{2}{*}{0693010401} & A* & all\\
& & B & all\\[0.2cm]

\multirow{3}{*}{ HD 190360} &  0304201101 & A* & all\\
& \multirow{2}{*}{0304202601} & A* & all\\
&&B&pn\\[0.2cm]

\multirow{2}{*}{Kepler-1008} & \multirow{2}{*}{0550451901} & A* & MOS2 \& pn \\
& & B & MOS2 \& pn \\[0.2cm]

\multirow{2}{*}{$\upsilon$ And}&\multirow{2}{*}{0722030101} &A*& all\\
&&B&all\\[0.2cm]

\multirow{2}{*}{WASP-18} & \multirow{2}{*}{0673740101} & A* & all \\
 & & B & all\\[0.2cm]

\multirow{2}{*}{WASP-33} & \multirow{2}{*}{0785120201} & A*B & MOS1 \& pn \\
& & C & pn \\[0.2cm]

\multirow{4}{*}{XO-2} & \multirow{2}{*}{0728970101} & S* & all\\
&&N*&all\\
& \multirow{2}{*}{0728970201} & S* & all\\
&&N*&all\\[0.2cm]
\hline
\end{tabular}
\caption{Systems observed by the XMM-Newton telescope. The components with the asterisk symbol are the planet-hosting stars. Most of the stellar components were properly positioned on a CCD chip in all three cameras (marked as {\it all} in the {\it camera} column), while in some observations a component was on the edge of the CCD array, or fell on a gap between chips. In these cases, the observation with the given camera was discarded.}
\label{table:1}
\end{table}

We extracted X-ray photon events from the source and background regions, restricting ourselves to events with the highest probability of being single-photon events as recommended in the \href{https://xmm-tools.cosmos.esa.int/external/xmm_user_support/documentation/uhb/index.html}{XMM-Newton data analysis handbook}.

The stellar coronae of cool stars are best observed in the 0.2-2.0 keV energy band since the bulk of magnetically induced coronal high energy radiation is emitted in this energy range (see for example \cite{Guedel1997}).
Therefore, we constrained our analysis to the source and background photon events within these energy values. For photon energies higher than 2 keV, usually, the background radiation dominates, lowering the signal-to-noise ratio for the source.

\begin{figure}
  \includegraphics[width=\linewidth]{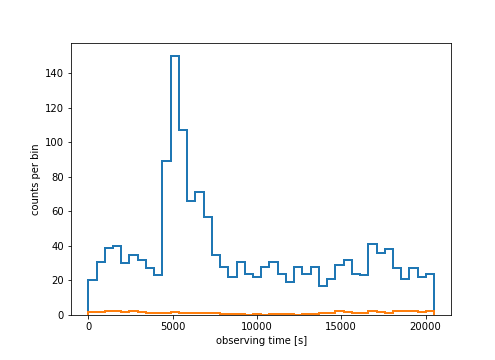}
  \caption{The light curve of the planet host GJ-15 A. The blue step function represents the photon counts in a 500s time interval, while the orange line at the bottom represents the scaled background level with the steps ranging the same time interval. The background extraction region is $\approx$ 14 times larger than the source extraction region of the planet host. Therefore, the background counts are scaled down by a factor of 14.}
  \label{fig:GJ15_flare}
\end{figure}

We defined Good Time Intervals (GTI) for each observation, i.e.\ we excluded times of high background signal (see Appendix A. for more details). Furthermore, cool stars are known to flare stochastically  \citep{Audard2000,Walkowicz2011, Ilin2019}. Since we wanted to avoid a flare dominating the X-ray brightness measurement of our stars and therefore skewing our activity level comparison, we inspected X-ray light curves for possible flaring events.
We found only one flaring event in the binary system GJ-15, occurring on the planet host (Fig. \ref{fig:GJ15_flare}). In the first half of the observation, a sharp increase in the source photon counts is observed, while the background level stays constant, followed by a gradual photon-count decrease. Therefore, we excluded this time interval from further analysis of this system and only used the quiescent time periods.

\subsubsection{Determining the excess source photons}
\label{KBN}

After acquiring the appropriate photon event lists, we had to estimate the number of photon events coming from the source itself. This we did by removing the background contribution from photon events gathered in the source extraction region. Hence, we estimated the {\it net source photon count (n)} and its confidence interval. Since both the source and background counts have statistical uncertainties following a Poisson process, suitable methods need to be used (see for example \cite{Ayres2004, KBN1991}).

In short, whenever we had a strong X-ray source that is visible by eye with more than 100 excess source counts over the background, we approximated the Poissonian uncertainties as Gaussian uncertainties and calculated the uncertainty on the net source counts {\it n} as the square root of the observed source counts {\it N}, thereby having a 68\% confidence interval of $CI = (n-\sqrt{N},n+\sqrt{N})$. In any case, for those bright sources, the intrinsic astrophysical variability of the star is expected to dominate over the statistical uncertainty of the net source counts.

For fainter sources, both the Poissonian uncertainties on the source and the background need to be taken into account. First, we estimated the probability that the background fluctuation was responsible for the number of counts in the source region by employing the Poisson cumulative distribution function. If the probability was higher than 0.3\%, we report an upper limit for the source, and otherwise, we report a detection with a 3-sigma level of detection significance.

To estimate the net source photon count and its confidence interval for a detected faint source, we applied the Kraft-Burrows-Nousek estimator \citep{KBN1991}. For faint sources over some background, it can in principle occur that the source region has fewer counts than the scaled background region, but this does not indicate negative source fluxes but is a result of small number statistics. The KBN estimator tackles this in a Bayesian manner by explicitly assuming the background signal to stem from a Poisson process as well, and marginalizing over all possible background count numbers in the source detect cell. It assumes the source flux to be non-negative, and yields confidence intervals for the net source count rate. We used the KBN implementation of the \texttt{stats.poisson\_conf\_interval} function in the \texttt{astropy} package in our analysis \citep{Astropy2013,Astropy2018}.

For the detected faint sources, we determine the 68\% confidence interval and report its center as the net source photon count for this detection. For undetected faint sources, we calculated the maximum X-ray luminosity by setting the upper end of the 95\% confidence interval to be the upper limit to the net source photon count.

\subsubsection{From X-ray counts to X-ray fluxes}
\label{conversionfactor}

To convert the estimated net source photon count into X-ray flux, we need to know the underlying X-ray spectrum of the observed stellar corona and the telescope's instrumentation setup during the observation. For bright sources, this could be done by spectral fitting, but for weaker sources, other inferences on the main variable, namely the coronal temperature, need to be made.
Regarding observations acquired with EPIC, we have an instrumentation setup that includes filters of different thicknesses (thin, medium, thick) and two camera types (pn and MOS). For each setup, there is a {\it photon count conversion factor (C)} as an estimate for the number of photons to be detected for a given source flux, mean coronal temperature, and energy range.

To estimate the mean coronal temperature of our sources, we calculated the hardness ratio of the detected radiation by utilizing the observations made with the PN camera. The hardness ratio represents the normalized difference of photon counts in two wavelength bands: HR = H-S/H+S. The wavelength bands we have used are  S = 0.2-0.7 keV and H = 0.7-2.0 keV \footnote{he division into the two passbands follows the example of the solar spectrum (\cite{Guedel1997}, Fig. 5): the photon count rate increases gradually with photon energy, reaching the maximum emissivity at around 0.7 keV and then decreasing steeply. Therefore, the division point was set to be at the peak of the solar X-ray emissivity.}. 
If $HR \gt 0$, we have a harder X-ray spectrum and a hotter stellar corona, while if $HR \lt 0$, the spectrum is softer and the stellar corona is cooler.

To calculate the conversion factors, we have used the NASA HEASARC online tool \href{https://heasarc.gsfc.nasa.gov/cgi-bin/Tools/w3pimms/w3pimms.pl}{WebPimms} \citep{Mukai1993}, version 4.11. With this tool, we estimated the hardness ratio of a hypothetical stellar corona (spectral model of hot plasma -- APEC) for a given mean coronal temperature in the energy range of 0.2-2.0 keV, with the source flux of $1e-12 \frac{erg}{cm^{2}s}$ and the metallicity of 1 solar abundance. No absorption due to interstellar matter was assumed since most of our X-ray detected sources are within 200\,pc from the Sun. Comparing a set of estimated hardness ratios for a given instrumental setup to the observed HR value, we determined the mean coronal temperature of our source of interest. If a source is marked as not detected in the soft or hard energy band, we assume a mean coronal temperature of 3 MK ($\log_{10}T = 6.477$) since this is the expected coronal temperature for a moderately active star \citep{Withbroe1977,Schmitt1990,Foster2021}. The conversion factors that were used for the conversion of photon counts into X-ray flux of XMM-Newton sources are given in table \ref{conversionfactorsXMM}.



Since the observed stellar components are detected as point sources, we also have to take into account the influence of the point spread function (PSF), which describes the dispersion of incoming photons onto the CCD due to their interaction with the telescope's optics. For this task, we used appropriate Encircled Energy Fractions (EEFs) for our photon extraction radii according to \href{https://xmmweb.esac.esa.int/docs/documents/CAL-TN-0029-1-0.pdf}{calibration documentation} for the in-flight calibration of the PSF for the PN camera. The EEF value typically lies between 0.4 and 0.7 and has a weak dependence on the optical off-axis angle.

Having all necessary parameters, including the observation exposure time, that can be extracted from the observation file header, the source X-ray flux can be calculated as the sum of the fluxes detected by all three cameras:
\begin{equation}
F = \sum_{i=MOS1,MOS2,pn} n_i \left(\sum_{i} t_{i}C_{i}EEF_{i} \right)^{-1} \times 10^{-12} \frac{erg}{cm^2 s},
\end{equation}
where {\it t} represents the exposure time.

Also, some systems were observed multiple times. In these cases, we combined the observations to achieve a better X-ray flux accuracy. For them, the flux was calculated as follows:

\begin{equation}
F = \sum_{i,j}  n_{ij} \left(\sum_j\left(\sum_{i} t_{i}C_{i} EEF_{i}\right)_j \right)^{-1} \times 10^{-12} \frac{erg}{cm^2 s},
\end{equation}
where {\it j} is the summation over the multiple observations. Here, we calculated the exposure time-weighted average of the flux from several observations.

\begin{table*}
\begin{tabular}{lccccccc}
\hline
\hline
system & component & net photon counts &  HR & $\log_{10} T$ &  $F_x \times 10^{-16} \left[\frac{erg}{cm^{-2}s}\right]$ & r[\textit{pc}] & $L_x \left[\frac{erg}{s}\right]$ \\
\hline

\multirow{2}{*}{16 Cyg} & B* & $ 276.5 \pm 22.091 $ & -0.46 & 6.464 & $59.336 \pm 4.741 $  & \multirow{2}{*}{$21.139 \pm 0.015$} & $(3.234 \pm 0.258) \times 10^{26}$ \\
 & AC & $955.658 \pm 34.22$ & -0.466 & 6.461 & $ 204.468 \pm 7.321 $  & & $(1.115 \pm 0.040) \times 10^{27}$ \\[0.2cm]

\multirow{2}{*}{30 Ari} & B*C & $21499.292 \pm 146.727$ & 0.111 & 6.665 & $12978.979 \pm 88.578$ & \multirow{2}{*}{$45.137 \pm 0.149$} & $(3.225 \pm 0.031 ) \times 10^{29}$ \\
 & A & $12657.131 \pm 112.606$ & -0.068 & 6.605 & $8104.375 \pm 72.101$ & & $(2.014 \pm 0.022) \times 10^{29}$ \\[0.2cm]

\multirow{2}{*}{55 Cnc} & A* & $228.903 \pm 16.031$ & -0.307 & 6.477 & $360.448 \pm 25.244$ & \multirow{2}{*}{$12.586 \pm 0.012$} & $(6.964 \pm 0.488)\times 10^{26}$ \\
 & B & $51.903_{-8.628}^{+9.297}$ & / & 6.477 & $81.729_{-13.586}^{+14.639}$  & & $(1.579 \pm 0.273) \times 10^{26}$ \\[0.2cm]

\multirow{2}{*}{83 Leo} & A (G III) & $2155.188 \pm 46.723$ & -0.056 & 6.564 & $1706.574 \pm 36.997$ & \multirow{2}{*}{$18.204 \pm 0.058$} & $(6.898 \pm 0.156) \times 10^{27}$ \\
 & B* & $160.2 \pm 13.342$ & -0.497 & 6.402 & $167.99 \pm 13.99$ & & $(6.791 \pm 0.567) \times 10^{26}$ \\[0.2cm]

AS 205 & A*B & $78.445_{-10.414}^{+11.068}$ & / & 6.477 & $102.988_{-13.673}^{+14.530}$ & $127.492 \pm 1.6$ & $(2.042 \pm 0.284) \times 10^{28}$  \\[0.2cm]

\multirow{2}{*}{GJ-15} & A* & $1465.231 \pm 39.090$ & -0.468 & 6.460 & $1052.417 \pm 28.076$ & \multirow{2}{*}{$3.5623 \pm 0.0005$} & $(1.629 \pm 0.043) \times 10^{26}$ \\
 & B & $398.089 \pm 21.679$ & -0.608 & 6.396 & $250.571 \pm 13.646 $ &  & $(3.879 \pm 0.211) \times 10^{25}$ \\[0.2cm]

\multirow{2}{*}{HAT-P-16} & A*B & $16.518_{-5.256}^{+5.926}$ & / & 6.477 & $ 17.190_{-5.470}^{+6.168}$ & \multirow{2}{*}{$226.637 \pm 4.388$} & $(1.077 \pm 0.367) \times 10^{28}$\\
 & C & $9.611_{-3.813}^{+4.485}$&  & 6.477 & $12.395_{-4.917}^{+5.784}$ & & $(7.766 \pm 3.366) \times 10^{27}$ \\[0.2cm]

\multirow{2}{*}{HD 27442} & A*(K III) & $152.68 \pm 15.906$ & -0.676 & 6.331 & $81.838 \pm 8.526$ & \multirow{2}{*}{$18.27 \pm 0.054$} & $(3.359 \pm 0.350) \times 10^{26}$ \\
 & B(WD) & $62.354_{-12.341}^{+12.991}$ & / & 6.477 & $31.171_{-6.169}^{+6.494}$ & & $(1.280 \pm 0.260) \times 10^{26}$ \\[0.2cm]

HD 46375 & A*B & $66.051_{-8.74}^{+9.407}$ & -0.561 & 6.418 & $110.982_{-14.684}^{+15.808}$ & $29.553 \pm 0.038$ & $(1.182 \pm 0.162) \times 10^{27}$ \\[0.2cm]

\multirow{2}{*}{HD 75289} & A* & $ 16.741_{-5.165}^{+5.836}$ & / & 6.477 & $11.923_{-3.679}^{+4.157}$ & \multirow{2}{*}{$29.116 \pm 0.024$} & $(1.233 \pm 0.405) \times 10^{26}$ \\
 & B & $235.682 \pm 16.062$ & -0.3 & 6.527 & $142.426 \pm 9.707$ & & $(1.473 \pm 0.100) \times 10^{27}$ \\[0.2cm]

\multirow{2}{*}{HD 101930} & A* & $ \le 30.94$ & / & 6.477 & $ \le 83.215 $ & \multirow{2}{*}{$30.027 \pm 0.041$} & $ \le 9.152 \times 10^{26}$ \\
 & B & $ \le 18.73$ & / & 6.477 & $ \le 61.519 $ &  & $ \le 6.766 \times 10^{26}$  \\[0.2cm]

\multirow{2}{*}{HD 107148} & A* & $32.125_{-6.121}^{+13.914}$ & / & 6.477 &  $20.281_{-5.503}^{+5.883}$ & \multirow{2}{*}{$49.416 \pm 0.116$} & $(6.041 \pm 0.17)\times 10^{26} $ \\
& B(WD) & $ \le 13.982 $ & / & 6.477 & $ \le 7.946 $ &  & $ \le 2.367 \times 10^{26}$ \\[0.2cm]

\multirow{2}{*}{HD 190360} & A* & $16.75_{-6.16}^{+6.829}$ & / & 6.477 & $28.143_{-10.351}^{+11.474}$ & \multirow{2}{*}{$16.007 \pm 0.016$} & $(8.796 \pm 3.410) \times 10^{25}$\\
 & B & $15.875_{-3.816}^{+4.49}$ & -0.449 & 6.469 & $116.032_{-27.892}^{+32.815}$ & & $(3.626 \pm 0.949) \times 10^{26}$ \\[0.2cm]

\multirow{2}{*}{Kepler-1008} & A* & $ \le 7.593 $ & / & 6.477 & $ \le 7.473 $ & \multirow{2}{*}{$282.565 \pm 1.278$} & $ \le 7.278\times10^{27}$ \\
& B & $ \le 10.541 $ & / & 6.477 & $ \le 10.374 $ &  & $ \le 1.01\times10^{28}$ \\[0.2cm]

\multirow{2}{*}{$\upsilon$ And} & A* &  $2286.302 \pm 48.61$ & -0.359 & 6.457 & $2740.445 \pm 58.266$ & \multirow{2}{*}{$13.405 \pm 0.063$} & $(6.007 \pm 0.140) \times 10^{27}$ \\
 & B & $58.314_{-10.639}^{+11.304}$ & / & 6.477  & $71.500_{-13.045}^{+13.861}$ & & $(1.567 \pm 0.295) \times 10^{26}$ \\[0.2cm]

\multirow{2}{*}{WASP-18} & A* & $ \le 13.857 $ & / & 6.477 & $ \le 31.465 $ & \multirow{2}{*}{$123.483 \pm 0.37$} & $ \le 5.852 \times 10^{27}$ \\
 & B & $ \le 11.578 $ & / & 6.477 & $ \le 26.290 $ & & $ \le 4.89 \times 10^{27}$ \\[0.2cm]

\multirow{2}{*}{WASP-33} & A*B & $ \le 26.869 $ & / & 6.477 & $ \le 53.839 $ & \multirow{2}{*}{$121.944 \pm 0.99$} & $ \le 9.766 \times 10^{27}$ \\
 & C & $40.0_{-8.11}^{+8.779}$ & 0.116 & 6.692 & $110.815_{-22.468}^{+24.322}$& & $(2.010 \pm 0.425) \times 10^{28}$ \\[0.2cm]

\multirow{2}{*}{XO-2} & S* & $78.137_{-17.103}^{+17.746}$ & / & 6.477 & $21.922_{-4.798}^{+4.979}$ & \multirow{2}{*}{$151.398 \pm 0.95$} & $(6.129 \pm 1.369) \times 10^{27}$\\
 & N* & $85.13_{-17.296}^{+17.953}$ & / & 6.477 & $23.884_{-4.852}^{+5.037}$ & & $(6.678 \pm 1.385) \times 10^{27}$ \\[0.2cm]

\hline
\end{tabular}
\caption{The parameters for evaluating the X-ray luminosity of binary companions observed with the XMM-Newton telescope are given. The asterisk symbol marks the planet-hosting component. If excess counts do not have a confidence interval given, only the upper limit of the X-ray flux has been estimated. Otherwise, if the confidence interval is symmetric/asymmetric, indicates that we have applied 'bright'/'faint'-source statistics (see chapter \ref{KBN}).}
\label{tab:XMM_Newton_flux}
\end{table*}


\subsection{Analysis of Chandra data}

\subsubsection{Chandra instrumentaion}

The Chandra X-ray Observatory has two focal plane science instruments, ACIS and HRC (High-Resolution Camera), which provide information about the number, position, energy, and arrival time of incoming X-ray photons.
The Chandra Advanced CCD Imaging Spectrometer (ACIS) is the only instrument employed for observing our sources of interest since it provides images as well as energy information about the detected photons. It is sensitive to the energy range of 0.2-10.0 keV and the on-axis angular resolution (FWHM of the PSF) is $\approx$ 0.5\arcsec at 0.277 keV. 

The ACIS instrument is composed of ten CCD chips in 2 arrays that provide imaging and spectroscopy: four CCDs (I0 to I3) are employed in the imaging array (ACIS-I), while the other six (S0-S5) compose the spectroscopic array (ACIS-S). The S1 and S3 CCDs are back-illuminated chips, meaning they have a higher sensitivity to X-ray photons. The FoV of ACIS-I is square-shaped measuring 16\arcmin $\times$16\arcmin, while the spectroscopic array of ACIS-S is elongated measuring an FoV of 8\arcmin $\times$48\arcmin.

The shape and size of the telescope PSF vary significantly with the source location on the detector and its spectral energy distribution. It provides the best resolution in an area centered about the optical axis, and it deteriorates strongly with increasing off-axis angle, letting the source appear as extended. Therefore, we took into account the source position on the detector when defining the size of the source extraction region.

Table \ref{table:3} lists all systems obtained in the data archive of the Chandra X-ray observatory.

\begin{table}
\centering
\begin{tabular}{|l|c|c|c|c|}
\hline
\hline
system & obs ID & star & chip & off-axis angle $[\arcsec]$ \\
\hline

\multirow{6}{*}{16 Cyg} & \multirow{3}{*}{16647} & A & \multirow{6}{*}{I3} & 28.9\\
&  & B* & & 25.0\\
&  & C & & 30.7\\
& \multirow{3}{*}{18756} & A & & 28.9\\
&  & B* & & 25.0\\
&  & C & & 30.7\\[0.2cm]

\multirow{2}{*}{55 Cnc} & 14401 & \multirow{2}{*}{A*} & \multirow{2}{*}{S3} & 0.7\\
& 14402 &  & & 1.0\\[0.2cm]

AS 205 & 16327 & A*B & S3 & 1.2\\[0.2cm]

\multirow{2}{*}{CoRoT-2} & \multirow{2}{*}{10989} & A* & \multirow{2}{*}{S3} & 7.5\\
& & B & & 10.5\\[0.2cm]

\multirow{2}{*}{HAT-P-20} & \multirow{2}{*}{15711} & A* & \multirow{2}{*}{S3} & 1.5\\
& & B & & 5.8\\[0.2cm]

\multirow{2}{*}{HAT-P-22} & \multirow{2}{*}{15105} & A* & \multirow{2}{*}{I3} & 258.8\\
& & B & & 267.1\\[0.2cm]

\multirow{2}{*}{HATS-65} & 3282 & \multirow{2}{*}{A*B} & \multirow{2}{*}{I2} & 592.8\\
& 9382 &  & & 592.9\\[0.2cm]

HIP 116454 & 19517 & A*B & S2 & 1070.0\\[0.2cm]

\multirow{2}{*}{HD 46375} & \multirow{2}{*}{15719} & A* & \multirow{2}{*}{S3} & 0.6\\
& & B & & 9.9 \\[0.2cm]

HD 96167 & 5817 & A*B & I3 & 505.5\\[0.2cm]

\multirow{2}{*}{HD 107148} & \multirow{2}{*}{13665} & A* & \multirow{2}{*}{S3} & 1.3 \\
& & B & & 35.7 \\[0.2cm]

\multirow{2}{*}{HD 109749} & \multirow{2}{*}{15720} & A* &\multirow{2}{*}{S3} & 3.1 \\
& & B & & 9.5\\[0.2cm]

\multirow{2}{*}{HD 178911} & \multirow{2}{*}{13659} & B* & \multirow{2}{*}{S3} & 2.1\\
& & AC & & 20.7\\[0.2cm]

\multirow{2}{*}{HD 185269} & \multirow{2}{*}{15721} & A & \multirow{2}{*}{S3} & 1.3\\
&&Bab&&3.7\\[0.2cm]

\multirow{2}{*}{HD 188015}&\multirow{2}{*}{13667} &A*& \multirow{2}{*}{S3} & 2.2\\
&&B&& 16.1\\[0.2cm]

\multirow{2}{*}{HD 189733} & \multirow{2}{*}{12340/5} & A* & \multirow{2}{*}{S3} & 0.5\\
& & B & & 12.2\\[0.2cm]

\multirow{4}{*}{HD 197037} & 7444 & \multirow{4}{*}{A*B} & I3 & 542.2\\
& 8598 &  & I2 & 542.2\\
& 9770 &  & I2 & 542.3\\
& 9771 &  & I2 & 542.4\\[0.2cm]

\multirow{2}{*}{Kepler-444} & 17733 & \multirow{2}{*}{A*BC} & \multirow{2}{*}{S3} & 402.0\\
& 20066 &  & & 9.9 \\[0.2cm]

$\upsilon$ And & 10976/9 & A* & S3 &4.2 \\[0.2cm]

\multirow{2}{*}{WASP-8} & \multirow{2}{*}{15712} & A* &\multirow{2}{*}{S3}& 1.6\\
&&B&&5.0\\[0.2cm]

WASP-18 & 14566 & A* & I3 & 1.1\\[0.2cm]

\multirow{2}{*}{WASP-77} & \multirow{2}{*}{15709} & A* & \multirow{2}{*}{S3} & 1.9\\
&&B& &4.4\\[0.2cm]

\hline
\end{tabular}
\caption{Given are the binary systems observed with the Chandra X-ray observatory. The asterisk symbol marks the planet-hosting companion. The chip column gives the CCD where the system was observed.}
\label{table:3}
\end{table}

\subsubsection{Photon event extraction}

The Chandra telescope archival data is calibrated, so no preprocessing, as was the case with XMM-Newton observation, was needed. We used the Chandra Interactive Analysis of Observations (CIAO) X-ray data analysis software, version 4.12, to extract the photon events of interest. First, we defined the source and background extraction regions around the objects of interest and source-free area, respectively. The radius of the source extraction region was between 1\arcsec and 2\arcsec, depending on the angular separation between the stellar components, while the background region had a radius of 15\arcsec. For sources with an off-axis angle $\gt 2\arcmin$, the source and background extraction regions were chosen to be proportionally larger with the distance to the detector's center. Also, some systems had a large off-axis angle ($\gt 9\arcmin$) and small angular separation between the components, making them appear unresolved. For these cases, we estimated the net source photon count and calculated the flux for the system as a whole.

Having an angular resolution of $0.5\arcsec$, the components of our systems were in most cases resolved. The downside of this resolving power is the lower effective area in comparison with XMM-Newton and therefore the smaller number of event counts we collected within an extraction region. As a result, the event counts were low in most cases.

Having the photon event lists for the sources and the background, we applied the same analysis method to determine the net source photon count and confidence interval as for the sources observed with the XMM-Newton telescope.

The same methodology for deriving flux conversion factors was applied here as well. Since Chandra's effective area has changed significantly over the years due to the accumulation of a contaminant of the optical blocking filters of ACIS\footnote{More details can be found in the \href{https://cxc.harvard.edu/ciao/why/acisqecontamN0013.html}{Chandra X-ray Observatory handbook.}}, we also calculated different conversion factors for the different observatory cycles. The results are listed in table \ref{conversionfactorsChandra}.

\begin{table*}
\begin{tabular}{lccccccccc}
\hline
\hline
system & component & net photon counts & HR & $\log_{10}T$ & $Fx \times 10^{-16} \left[\frac{erg}{cm^2 s}\right]$  & r[pc] & $L_x \left[\frac{erg}{s}\right]$ \\
\hline

\multirow{3}{*}{16 Cyg} & B* & $ \le 8.786 $& / & 6.477 & $ \le 37.352 $ & \multirow{3}{*}{$21.14 \pm 0.011$} & $ \le 2.036 \times 10^{26}$ \\
 & A  & $42.543_{-6.243}^{+6.914}$ & 0.863 & 6.717 & $40.930_{-6.007}^{+6.651}$ & & $(2.231 \pm 0.345) \times 10^{26}$ \\
 & C & $4.718_{-1.944}^{+2.632}$ & / & 6.477 & $20.055_{-8.263}^{+11.190}$ & & $(1.093 \pm 0.53) \times 10^{26}$ \\[0.2cm]

55 Cnc & A* & $42.235_{-6.244}^{+6.913}$ & -0.865 & 6.226 & $55.020_{-8.134}^{+9.007}$ & $12.586 \pm 0.012$ & $(1.063 \pm 0.166) \times 10^{26}$ \\[0.2cm]

AS 205 & A*B & $48.935_{-6.686}^{+7.355}$ & 0.918 & 7.18 & $657.214_{-89.790}^{+98.784}$ & $127.492 \pm 1.6$ & $(1.303 \pm 0.19) \times 10^{29}$ \\[0.2cm]

\multirow{2}{*}{CoRoT-2} & A* & $62.000_{-8.051}^{+8.72}$ & / & 6.934 & $210.000_{-130.000}^{+80.000}$& \multirow{2}{*}{$213.283 \pm 2.457$} & $(1.165 \pm 0.583) \times 10^{29}$ \\
 & B & $ \le 3.309 $ & / & 6.477 & $ \le 25.471 $ & & $ \le 1.413 \times 10^{28}$ \\[0.2cm]

\multirow{2}{*}{HAT-P-20} & A* & $25.920_{-4.788}^{+5.460}$ & 0.618 & 6.711 & $140.492_{-25.954}^{+29.594}$ & \multirow{2}{*}{$71.037 \pm 0.199$} & $(8.648 \pm 1.71) \times 10^{27}$ \\
 & B & $1.928_{-1.136}^{+1.856}$ & / & 6.477 & $14.242_{-8.393}^{+13.708}$ & & $(8.767 \pm 6.802) \times 10^{26}$ \\

\multirow{2}{*}{HAT-P-22} & A* & $ \le 4.238 $ & / & 6.477 & $ \le 63.945 $ & \multirow{2}{*}{$81.765 \pm 0.25$} & $ \le 5.215 \times 10^{27}$ \\
 & B & $ \le 8.684 $ & / & 6.477 & $ \le 131.033 $ & & $ \le 1.069 \times 10^{28}$  \\[0.2cm]

HATS-65 & A*B & $ \le 22.508 $ & /& 6.477 & $ \le 32.045 $ & $493.32 \pm 12.317$ & $ \le 9.513 \times 10^{28}$\\[0.2cm]

HIP 116454 & A*B & $ \le 6.25 $ & / & 6.477 & $ \le 291.912 $ & $62.449 \pm 0.183$ & $ \le 1.389 \times 10^{28}$ \\[0.2cm]

\multirow{2}{*}{HD 46375} & A* & $16.989_{-3.816}^{+4.489}$ & -0.177 & 6.426 & $137.573_{-30.902}^{+36.357}$ & \multirow{2}{*}{$29.553 \pm 0.038$} & $(1.466 \pm 0.358) \times 10^{27}$\\
 & B & $5.989_{-2.155}^{+2.839}$ & -0.336 & 6.382 & $52.064_{-18.731}^{+24.685}$ & & $(5.547 \pm 2.313) \times 10^{26}$ \\[0.2cm]

HD 96167 & A*B & $5.032_{-2.154}^{+2.839}$ & / & 6.477 & $76.234_{-32.637}^{+43.011}$ & $85.301 \pm 0.414$ & $(6.766 \pm 3.358) \times 10^{27}$ \\[0.2cm]

\multirow{2}{*}{HD 107148} & A* & $4.93_{-1.944}^{+2.632}$ & -0.213 & 6.438 & $30.203_{-11.908}^{+16.125}$ & \multirow{2}{*}{$49.416 \pm 0.116$} & $(8.996 \pm 4.175) \times 10^{26}$ \\
 & B (WD) & $ \le 2.996$ & / & 6.477 & $ \le 17.414 $ & & $\le 5.187 \times 10^{26}$  \\[0.2cm]

\multirow{2}{*}{HD 109749} & A* & $14.907_{-3.567}^{+4.242}$  & -0.333 & 6.383 & $46.552_{-11.139}^{+13.245}$ & \multirow{2}{*}{$63.082 \pm 0.295$} & $(2.260 \pm 0.592) \times 10^{27}$ \\
 & B & $31.907_{-5.345}^{+6.016}$ & 0.252 & 6.563 & $75.514_{-12.649}^{+14.237}$ & & $(3.665 \pm 0.653) \times 10^{27}$ \\[0.2cm]

 \multirow{2}{*}{HD 178911} & B*  & $18.910_{-4.051}^{+4.724}$ & -0.155 & 6.455 & $109.073_{-23.365}^{+27.246}$ & \multirow{2}{*}{$49.394 \pm 0.946$} & $(3.246 \pm 0.763) \times 10^{27}$ \\
 & AC  & $64.910_{-7.746}^{+8.416}$ & 0.171 & 6.566 & $331.603_{-39.575}^{+42.993}$ & & $(9.869 \pm 1.285) \times 10^{27}$ \\[0.2cm]

\multirow{2}{*}{HD 185269} & A* (G IV) & $10.127_{-3.014}^{+3.691}$  & -0.277 & 6.398 & $16.552_{-4.926}^{+6.033}$ & \multirow{2}{*}{$51.992 \pm 0.084$} & $(5.458 \pm 1.807) \times 10^{26}$ \\
 & Bab & $8.603_{-2.7}^{+3.379}$  & 0.126 & 6.513 & $11.577_{-3.633}^{+4.547}$ & & $(3.817 \pm 1.349) \times 10^{26}$ \\[0.2cm]

\multirow{2}{*}{HD 188015} & A* & $16.968_{-3.816}^{+4.489}$ & -0.175 & 6.449 & $75.860_{-17.061}^{+20.073}$ & \multirow{2}{*}{$50.671 \pm 0.109$} & $(2.376 \pm 0.582) \times 10^{27}$ \\
 & B & $7.986_{-2.53}^{+3.21}$ & / & 6.477 & $34.383_{-10.891}^{+13.822}$ & & $(1.077 \pm 0.387) \times 10^{27}$ \\[0.2cm]

\multirow{2}{*}{HD 189733} & A* & $6525.709 \pm 80.796$ & 0.258 & 6.6 & $2720.596 \pm 33.684$ & \multirow{2}{*}{$19.764 \pm 0.013$} & $(1.296 \pm 0.016) \times 10^{28}$ \\
 & B & $364.813 \pm 19.131$ & 0.187 & 6.573 & $155.993 \pm 8.180$ & & $(7.433 \pm 0.390) \times 10^{26}$ \\[0.2cm]

HD 197037 & A*B & $27.39_{-6.244}^{+6.913}$ & 0.378 & 6.491 & $9.840_{-2.243}^{+2.484}$ & $33.194 \pm 0.029$ & $(1.322 \pm 0.318) \times 10^{26}$ \\[0.2cm]

Kepler-444 & A*BC & $4.362_{-1.944}^{+2.632}$ & / & 6.477 & $13.164_{-5.866}^{+7.943}$ & $36.44 \pm 0.039$ & $(2.132 \pm 1.118) \times 10^{26}$\\[0.2cm]

$\upsilon$ And & A* & $2836.45 \pm 53.272$  & -0.171 & 6.481 & $2266.694 \pm 42.572$ & $13.405 \pm 0.063$ & $(4.969 \pm 0.104) \times 10^{27}$ \\[0.2cm]

\multirow{2}{*}{WASP-8} & A*  & $42.955_{-6.244}^{+6.914}$  & 0.21 & 6.579 & $218.160_{-31.710}^{+35.113}$ & \multirow{2}{*}{$89.961 \pm 0.363$} & $(2.154 \pm 0.33) \times 10^{28}$ \\
 & B  & $ \le 2.996 $ & / & 6.477 & $ \le 16.907 $ & & $\le 1.669 \times 10^{27}$ \\[0.2cm]

WASP-18 & A* & $ \le 4.228 $ & / & 6.477 & $ \le 10.384 $ & $123.483 \pm 0.37$ & $ \le 1.931 \times 10^{27}$ \\[0.2cm]

\multirow{2}{*}{WASP-77} & A* & $24.899_{-4.69}^{+5.361}$ & 0.363 & 6.608 & $154.996_{-29.194}^{+33.375}$ & \multirow{2}{*}{$105.166 \pm 1.196$} & $(2.091 \pm 0.425) \times 10^{28}$ \\
 & B & $ \le 4.648 $ & / & 6.477 & $ \le 34.336 $ & & $ \le 4.632 \times 10^{27}$ \\[0.2cm]

\hline
\end{tabular}
\caption{The parameters for evaluating the X-ray luminosity of binary companions observed with the Chandra space telescope are given. The asterisk symbol marks the planet-hosting component. The differentiation between non-detections and 'faint'/'bright'-source statistics is the same as in Table \ref{tab:XMM_Newton_flux}. Here we note that the given coronal effective temperature and the X-ray flux for CoRoT-2 A are adopted from \protect\cite{Schroter2011}, while the X-ray luminosity was calculated using the geometric distance given by \protect\cite{BailerJones2018}. Additionally, the CoRoT-2 A X-ray flux uncertainty corresponds to the limits of a 90\% confidence interval (as given in \protect\cite{Schroter2011}), while the flux uncertainty given for other sources of our sample corresponds to the 68\% confidence interval.}
\label{tab:Chandra_flux}
\end{table*}




\subsection{X-ray Luminosity}

As the last step in the data analysis process, we calculated the X-ray luminosity of our sources. For this, we used the distances calculated by \cite{BailerJones2018}. They used a Bayesian method to infer geometric distances from parallax measurements done by the ESO Gaia mission provided in the second data release (DR2). For all stellar components of a system, we used the distance given for the primary component. We calculated the luminosity uncertainty by using the 1-sigma flux uncertainty and the distance uncertainty and propagating the uncertainties. The net source photon counts, X-ray flux, and luminosity for the system components observed with the XMM-Newton and Chandra telescopes are given in Tables \ref{tab:XMM_Newton_flux} and \ref{tab:Chandra_flux}, respectively. Additionally, parameters calculated as intermediate steps are given. Here, we want to point out that the coronal temperature and the X-ray flux of CoRoT-2 A were adopted from \cite{Schroter2011}. The given flux uncertainties represent the 90\% confidence interval. Its X-ray luminosity was calculated with the distance given in Table \ref{tab:Chandra_flux}. For more details on the analysis of the CoRoT-2 system see Section \ref{ch:corot-2}.


\section{Results}
\label{results}
\subsection{Mass estimates of the sample stars}
\label{mass}
Stars of different masses have different spin-down time scales and therefore different activity levels at a given stellar age. To make the activity levels of the binary companions comparable to one another, we need the knowledge of their stellar spectral type and we chose the stellar mass as the spectral type indicator. For this task, we used the \textit{G-Rp} color and apparent \textit{G} magnitude of individual sources published in the second Gaia data release DR2 \citep{Gaia1,Gaia2}.
As described in \cite{Lindegren2018} and \cite{Arenou2018}, we corrected the parallax for the zero point parallax value, checked if the quality indicators 
are within recommended ranges, and corrected the value of the apparent magnitude\footnote{For the magnitude correction we also used the prescription given in the work of \cite{MaizApellaniz2018}.}. Additionally, we calculated the re-normalized unit weight error (RUWE) parameter as described in \cite{LL:LL-124} for all sources. 

We calculated the absolute magnitude assuming no absorption due to interstellar matter and, having the parallax and magnitude uncertainty, we calculated the absolute magnitude uncertainty using the error propagation function. The color and magnitudes for each star of our final sample, for which we estimated a mass, is given in Table \ref{tab:gaia_params}.
Some of our sources did not pass the Gaia quality assessment or were below the detection threshold or at the saturation level of the instruments. For these sources, we acquired the stellar mass parameter from the literature, where spectra were used for the spectral type determinations.

We employed an additional check on the stellar sample by testing whether the sample stars are sufficiently close to the main sequence. Stars younger than the main sequence have likely not experienced a significant accumulated tidal interaction with their planets, and stars older than the main sequence have marked changes in their rotation due to their growing radii, making direct activity comparisons inconclusive.

We, therefore, defined the Main Sequence (MS) in the Gaia CMD by using the stellar properties published by \cite{Pecaut2013} in their updated version available as an \href{https://www.pas.rochester.edu/~emamajek/EEM_dwarf_UBVIJHK_colors_Teff.txt}{online table}. We fitted a $7^{th}$ order polynomial to the $G-R_p$ color and the absolute G magnitude of MS stars, and defined a corridor of 0.1 mag at sunlike temperatures, widening linearly to 0.3 mag at M dwarf temperatures to capture the observed spread in absolute M dwarf brightnesses \citep{Kiman2019}.


Therefore, we were able to estimate by interpolation if a given source with observed Gaia color and magnitude falls, within the calculated uncertainties, on the canonical MS as defined by \cite{Pecaut2013}. If the stellar source is too faint in magnitude for observed color and too blue for observed magnitude, we mark it as being below the main sequence. Vice versa is true for sources marked as being above the main sequence. For MS stars, again by interpolation, the stellar mass was determined. The X-ray luminosity and mass values, together with the stellar spectral type, are given in Table \ref{tab:gaiamasses}.



\subsection{Activity level difference}

As previously discussed, the X-ray luminosity of the stellar companion of a planet-hosting star is used to assess whether the activity of the host star is typical or not, given that the components have co-evolved\footnote{See the work by \cite{White2001} and \cite{Kouwenhoven2010} for the discussion on the probability of co-evolution in multiple star systems.}. When the stellar components of a system have the same spectral type, the difference in the activity level can be determined by directly comparing the activity indicators. However, to make the X-ray luminosities of stellar components with different spectral types comparable to each other, we need to normalize their $L_x$ values, and for this task, we used the NEXXUS database of stellar X-ray emission \citep{SchmittLiefke2004}.

In the NEXXUS database, ROSAT observations were employed to estimate the X-ray luminosities of a volume-limited sample of F/G-, K-, and M-type field stars in the solar neighborhood. The detection completeness of the data set for F/G-type stars within 14 pc of the Sun is 94\%, for K-type stars within 12 pc is 96\%, and for M- and later type stars within 6 pc of the Sun is 91\%, where two of the non-detections are brown dwarf substellar objects. Therefore, it is assumed that the database sufficiently reproduces the X-ray luminosity distribution of stars in the solar neighborhood. With the advent of the eROSITA mission \citep{eROSITA}, there will be X-ray data of larger volume complete stellar samples available in the future, but as of now, the ROSAT data still represents the best volume-complete view of stellar coronae available.

\begin{figure}
  \includegraphics[width=\linewidth]{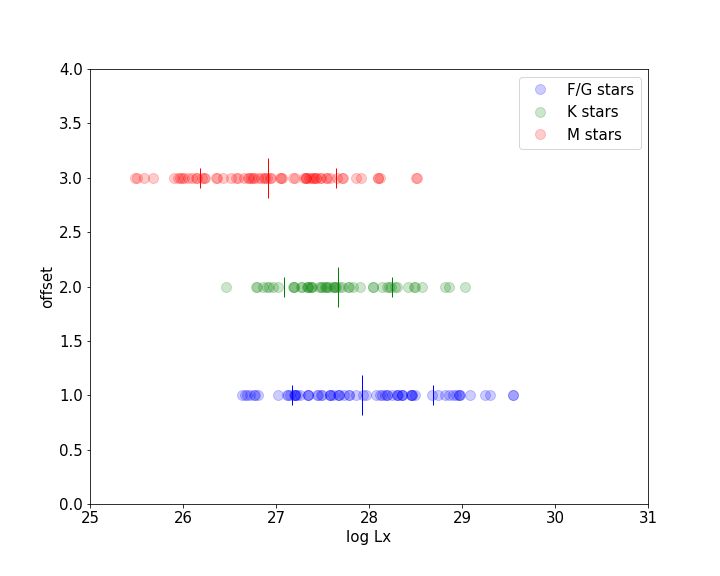}
  \caption{Shown is the X-ray luminosity distribution of stellar objects found in the NEXXUS database if they were to be observed in the 0.2-2.0 keV energy range. The distribution of M, K, and F/G type stars are represented by red, green, and blue dots, respectively. The mean of the distribution ($50^{th}$ percentile) is set by the long vertical line, while the standard deviation is set by the short vertical lines (the $15.87^{th}$ percentile on the left side of the mean or the $84.13^{th}$ percentile on the right side).}
  \label{fig:field_stars_dist}
\end{figure}

Two main conclusions from the analysis of the solar neighborhood sample were, that all main-sequence stars with outer convection zone are surrounded by hot coronae (I) and that the X-ray luminosities of cool dwarf stars extend over three orders of magnitude with the mean values decreasing
with decreasing spectral type (II) (see Fig. \ref{fig:field_stars_dist}). Our further analysis of the activity level difference was based on the second conclusion.

Since ROSAT observed in the energy range between 0.1 - 2.4 keV, the X-ray luminosities of solar neighborhood stars given in the NEXXUS database have a somewhat higher value than they have in the 0.2 - 2.0 keV energy range. \cite{Foster2021} calculated the conversion factor that has to be applied when converting from one energy range to the other:
\begin{equation}
\label{ROSAT_conversion}
F_{X,0.2 - 2.0 keV} = 0.87 \times F_{X,0.1 - 2.4 keV}.
\end{equation}

In Fig. \ref{fig:field_stars_dist}, we show the X-ray luminosity distributions of objects of different stellar spectral types analyzed in the solar neighborhood sample. The given X-ray luminosity values are corrected and represent the distribution if the stars have been observed in the 0.2-2.0 keV energy range. Although the distributions overlap in their luminosity values, we can estimate the mean value and the standard deviation of each one. If we have a system with stellar components of different spectral types with measured X-ray luminosities, we can use the mean value and standard deviation of the appropriate luminosity distributions and estimate at which percentile in the distribution the components are found. If both stars have a similar state of their rotation and activity evolution, they should be found at similar percentiles of their respective distribution.
Then, having the percentile value for each component, we can subtract them from each other and use the difference as an indicator for the discrepancy in the activity level of the stellar components.

Note that the ROSAT volume-complete sample consists of nearby field stars, which have a moderately wide range of ages around the age of our Sun. We will see in the following analysis that none of the stars of our wide binary sample fall outside the range of X-ray luminosities of the ROSAT sample, indicating that the age and activity ranges of our studied sample are reasonably well represented by the ROSAT sample.

\begin{figure*}
  \includegraphics[width=\textwidth]{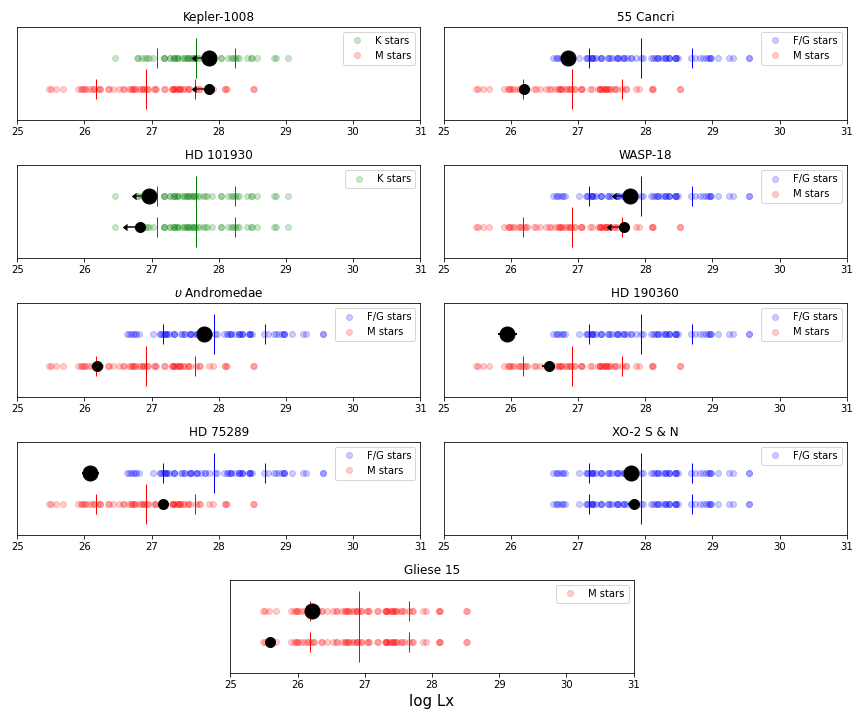}
  \caption{Shown are binary systems observed with the XMM-Newton space telescope. The systems shown here have stellar components that are on the Main Sequence. The colored dots represent the X-ray luminosity distribution of stars of the respective spectral type, as shown in Fig. \ref{fig:field_stars_dist}, while the large and the small black dots represent the primary and secondary stellar component, respectively.}
  \label{fig:XMM_Newton_systems}
\end{figure*}

\begin{figure*}
  \includegraphics[width=\textwidth]{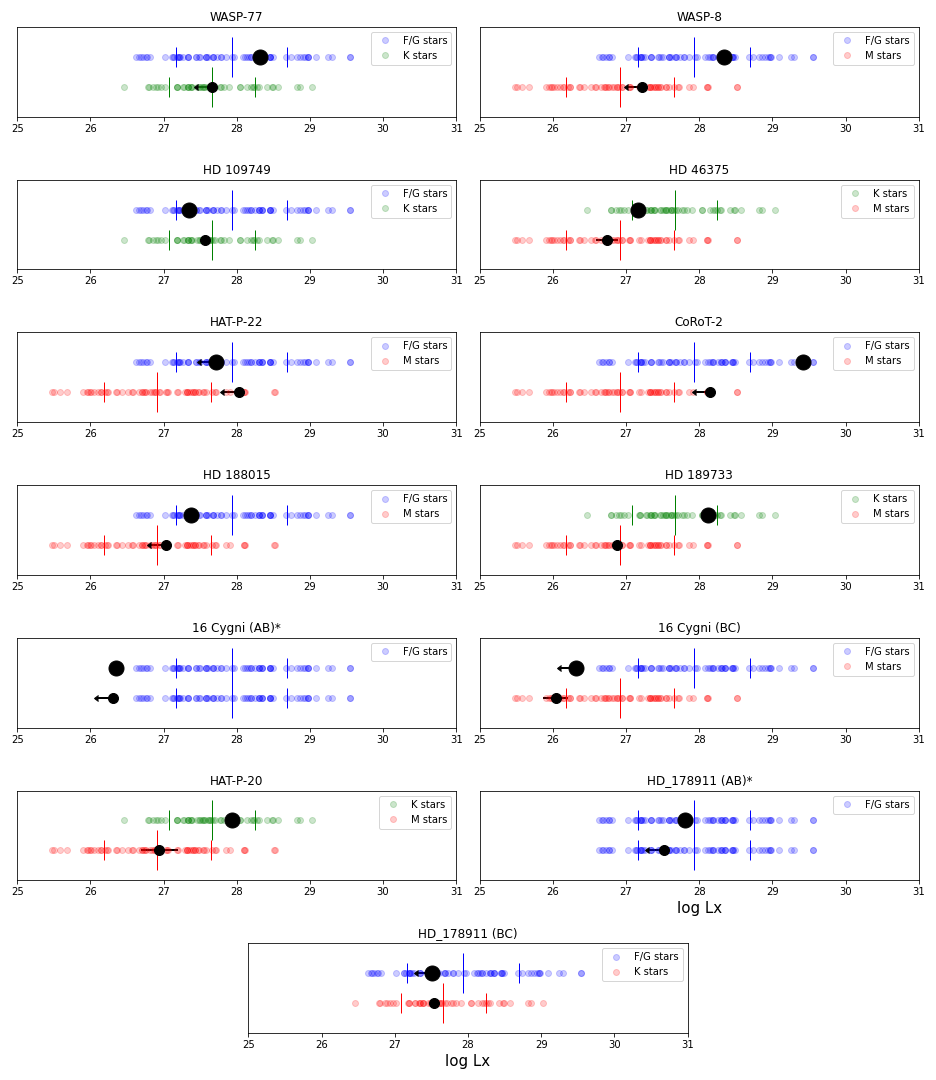}
  \caption{Shown are systems observed with the Chandra space observatory. The markings are the same as in Fig. \ref{fig:XMM_Newton_systems}. Only resolved systems with stellar components on the main sequence are represented. Since the Chandra space observatory has a better resolving power than the XMM-Newton space telescope, two systems that appeared unresolved with XMM-Newton (\textit{HD 46375} and \textit{16 Cyg}) are now resolved.}
  \label{fig:Chandra_systems}
\end{figure*}

In Fig. \ref{fig:XMM_Newton_systems}, we show some of the binary stellar systems that were observed with the XMM-Newton space telescope. Here, we did not consider systems that have evolved stellar components (HD 27442 and HD 107148) and where two components appear unresolved (HAT-P-16, 30 Ari, HD 46375, AS 205, 16 Cyg, and WASP-33). We also show systems where both stellar components are undetected. However, these system were not used in the final sample.

In Fig. \ref{fig:Chandra_systems} are shown some of the systems observed with the Chandra space observatory. From the initial data set, the evolved (HD 107148), unresolved (Kepler-444, HD 197037, HATS-65, AS 205), and evolved \& unresolved (HIP 116454, HD 185269, and HD 96167) systems are disregarded. Further, there were three systems (WASP-18, $\upsilon$ And, and 55 Cnc) that had only their primary component in the FoV of the camera, which also resulted in their exclusion. Also, the undetected systems HAT-P-22 was discarded from further analysis. Additionally, two of the three components of the system HD 178911 appeard unresolved which resultet in its initial exclusion; however, an analysis method was later found that enabled the inclusion of this system in the final sample. For more details, see Appendix \ref{appendA}.

Our final sample of stellar systems that were analyzed consists of 55~Cnc, GJ-15, HD~75289, HD~190360, $\upsilon$~And, and X0-2, which were observed by the XMM-Newton telescope and 16~Cyg, CoRoT-2, HAT-P-20, HD~46375, HD~109749, HD~178911, HD~188015, HD~189733, WASP-8, and WASP-77, that were observed with the Chandra space observatory.


\begin{table*}
\begin{tabular}{lccccccccc}
\hline
\hline
system & component & $L_x \left[\frac{erg}{s}\right]$ & Obs & Mass[$M_{\odot}$] & SpT & percentile value & Mass reference \\
\hline
\multirow{3}{*}{16 Cyg} & B* & $ \le 2.036 \times 10^{26}$ & \multirow{3}{*}{C} & 1.060 & G & $\le 0.0164$  & this work\\
 & A & $(2.231 \pm 0.345) \times 10^{26}$ & & $1.04^{+0.13}_{-0.12}$ & G & 0.0187 & \cite{TICv8}\\
 & C & $(1.093 \pm 0.53) \times 10^{26}$ & & / & M & 0.1148 & \cite{Patience2002} \\[0.2cm]
\multirow{2}{*}{55 Cnc} & A* & $(6.964 \pm 0.488)\times 10^{26}$ & \multirow{2}{*}{X}  & 0.909 & G & 0.0763 & \multirow{2}{*}{this work} \\
 & B & $(1.579 \pm 0.273) \times 10^{26}$ & & 0.237 & M & 0.163 & \\[0.2cm]
\multirow{2}{*}{CoRoT-2} & A* & $(1.165 \pm 0.583) \times 10^{29}$ & \multirow{2}{*}{C} & $0.97 \pm 0.14 ^a$ & G & 0.933 & \cite{TICv8}\\
 & B & $\le 1.413 \times 10^{28}$ & & 0.490 & M & $ \le 0.9547$ & this work\\[0.2cm]
\multirow{2}{*}{GJ-15} & A* & $(1.629 \pm 0.043) \times 10^{26}$ & \multirow{2}{*}{X} & 0.432 & M & 0.1674 & \multirow{2}{*}{this work}\\
 & B & $(3.879 \pm 0.211) \times 10^{25}$ & & 0.180 & M & 0.0345 & \\[0.2cm]
\multirow{2}{*}{HAT-P-20} & A* & $(8.648 \pm 1.71) \times 10^{27}$ & \multirow{2}{*}{C} & 0.702 & K & 0.681 & this work\\
 & B & $(8.767 \pm 6.802) \times 10^{26}$ & & 0.57 & M & 0.515 & \cite{Fontanive2019}\\[0.2cm]
\multirow{2}{*}{HD 46375} & A* & $(1.466 \pm 0.358) \times 10^{27}$ & \multirow{2}{*}{C} & 0.892 & K & 0.1939 & \multirow{2}{*}{this work} \\
 & B & $(5.547 \pm 2.313) \times 10^{26}$ & & 0.498 & M & 0.4072 & \\[0.2cm]
\multirow{2}{*}{HD 75289} & A* & $(1.233 \pm 0.405) \times 10^{26}$ & \multirow{2}{*}{X} & 1.183 & F & 0.0077 & this work \\
 & B & $(1.473 \pm 0.100) \times 10^{27}$ & & $0.14 \pm 0.02$ & M & 0.6355 & \cite{TICv8}\\[0.2cm]
\multirow{2}{*}{HD 109749} & A* & $(2.260 \pm 0.592) \times 10^{27}$ & \multirow{2}{*}{C} & 1.151 & F & 0.2244 & this work \\
 & B & $(3.665 \pm 0.653) \times 10^{27}$ & & 0.780 & K & 0.4307 & \cite{Desidera2007}\\[0.2cm]
\multirow{3}{*}{HD 178911} & B* & $(3.246 \pm 0.763) \times 10^{27}$ & \multirow{3}{*}{C} & 0.987 & G & 0.2911 & this work\\
 & A & $ \le 6.389 \times 10^{27}$ & & 1.1 & G & $\le 0.4354$ & \multirow{2}{*}{\cite{Tokovinin2000}} \\
 & C & $\le 3.479 \times 10^{27}$ & & 0.79 & K & $ \le 0.4154 $ & \\[0.2cm]
\multirow{2}{*}{HD 188015} & A* & $(2.376 \pm 0.582) \times 10^{27}$ & \multirow{2}{*}{C} & 1.038 & G & 0.2331 & \multirow{2}{*}{this work}\\
 & B & $(1.077 \pm 0.387) \times 10^{27}$ & & 0.180 & M & 0.5636 & \\[0.2cm]
\multirow{2}{*}{HD 189733} & A* & $(1.296 \pm 0.016) \times 10^{28}$ & \multirow{2}{*}{C} & 0.801 & K & 0.7808 & \multirow{2}{*}{this work}\\
 & B & $(7.433 \pm 0.390) \times 10^{26}$ &  & 0.193 & M & 0.4759 & \\[0.2cm]
\multirow{2}{*}{HD 190360} & A* & $(8.796 \pm 3.410) \times 10^{25}$ & \multirow{2}{*}{X} & 1.039 & G & 0.0045 & \multirow{2}{*}{this work} \\
 & B & $(3.626 \pm 0.949) \times 10^{26}$ & & 0.189 & M & 0.3129 &  \\[0.2cm]
\multirow{2}{*}{$\upsilon$ And} & A* & $(6.007 \pm 0.140) \times 10^{27}$ & \multirow{2}{*}{X} & 1.322 & F & 0.4216 & \multirow{2}{*}{this work} \\
 & B & $(1.567 \pm 0.295) \times 10^{26}$ & & 0.184 & M & 0.1619 & \\[0.2cm]
\multirow{2}{*}{WASP-8} & A* & $(2.154 \pm 0.33) \times 10^{28}$ & \multirow{2}{*}{C}  & 0.981 & G & 0.7028 & \multirow{2}{*}{this work} \\
 & B & $\le 1.669 \times 10^{27}$ & & 0.492 & M & $ \le 0.6631$ & \\[0.2cm]
\multirow{2}{*}{WASP-77} & A* & $(2.091 \pm 0.425) \times 10^{28}$ & \multirow{2}{*}{C} & 0.954 & G & 0.6969 & this work\\
 & B & $\le 4.632 \times 10^{27}$ & & / & K & $ \le 0.5006$ & \cite{Maxted2013}\\[0.2cm]
\multirow{2}{*}{XO-2} & S* & $(6.129 \pm 1.369) \times 10^{27}$ & \multirow{2}{*}{X} & 0.948 & G & 0.4261 & \multirow{2}{*}{this work} \\
 & N* & $(6.678 \pm 1.385) \times 10^{27}$ & & 0.947 & G & 0.4454 &  \\
\hline
\end{tabular}
\caption{Here, we show the calculated luminosities, masses, spectral types, and percentile values for each stellar component of systems that have been used in SPI modeling. The Obs column presents the observatory with which the data was acquired: {\bf X} denotes the XMM-Newton telescope, while {\bf C} stands for the Chandra observatory. The stellar companions 16 Cyg C and WASP-77 B had only their spectral type determined in the corresponding publications. $^a:$ By employing the method described in Section \ref{mass}, the estimated mass of CoRoT-2 A was 0.85 $M_{\odot}$, which would label it as a K-dwarf. Since this star is a known G-dwarf \protect\citep{Schroter2011,TICv8}, we discarded our calculated stellar mass. Here, we note that this was the only star for which the calculated mass and stellar mass from the TESS Input Catalog led to different spectral type estimates.}
\label{tab:gaiamasses}
\end{table*}

\subsubsection{The CoRoT-2 system}
\label{ch:corot-2}

The CoRoT-2 system takes a special place in the sample. It has a relatively large distance, making it possible that interstellar absorption influences the Chandra X-ray measurements. It has been investigated previously by \cite{Schroter2011}, who also performed a more detailed spectral analysis.
From their spectral analysis of the source photons of the primary, an interstellar absorption with a depth of the absorption column of $\approx 10^{21} cm^{-2}$ and a distance of $270 \pm 120$ pc was determined.
By using an APEC thermal model and fitting it to the X-ray spectrum of CoRoT-2 A, an X-ray luminosity of $1.9 \times 10^{29} erg/s$ in the 0.3-4.0 keV energy band was estimated. The secondary component, CoRoT-2 B, was not detected and an upper limit on the X-ray luminosity was determined: $L_X \lt 9 \times 10^{26}$. \cite{Schroter2011} also show that there is very little detectable stellar emission below 0.7 keV due to the ISM absorption.


While the updated stellar distances from \cite{BailerJones2018} place the system at a distance of $\approx 213$ pc instead of 270\,pc, the general analysis of \cite{Schroter2011} still applies and we use their coronal properties instead of the ones derived from a simple hardness ratio analysis.

However, when we analyse the secondary in the energy range of 0.7-2\,keV, we find that the KBN methodology gives a higher upper limit to the secondary's source flux than the one estimated by \cite{Schroter2011}. Specifically, we find that the 95\% upper limit to the secondary's count number is 3 photons, which translates into an upper limit of $L_x \lt 1.413 \times 10^{28} \frac{erg}{s}$. Adopting our estimate of the X-ray luminosity upper limit of the secondary places a more conservative estimate on the activity excess of the planet-hosting primary. Therefore, we used the spectral analysis from \cite{Schroter2011} for the primary and our luminosity estimate for the secondary in further analysis.

Furthermore, the available data from Gaia shows that the secondary is more likely to be an M dwarf instead of a late K dwarf, as was estimated by \cite{Schroter2011} and \cite{Evans2016}. Therefore, we continue our analysis of the CoRoT-2 system under the assumption that it is a G- and M-dwarf pair.


\subsection{Star-Planet Interaction Models}

The determined difference in activity levels of stars in a system can be related to expected star-planet interaction strength. Different models exist that explore potential tidal star-planet interactions with respect to different observable parameters. \cite{Miller2015}, for example, use the ratio of planet mass and the square of its semi-major axis to find systems for which star-planet interaction can be expected\footnote{The authors use the limit of $\log_{10} M_P/a^2 = 10 M_{Jup}AU^{-2}$ as the preferred limit for interaction strength between weakly and strongly interacting systems. According to the given limit, weakly interacting systems from our final sample are HD 190360 A,GJ 15 A, HD 188015 A, and 16 Cyg B.}. Although the authors were concerned with magnetic interaction, these two parameters are also applicable for the tidal interaction strength estimate. Other models additionally use stellar parameters like mass or radius to better define the level of tidal interaction between a planet and its host star. We have chosen three such models, where we used the derived tidal interaction parameter to compare to the measured activity level differences: The first model estimates the timescales for tidal dissipation due to SPI \citep{Albrecht2012}, the second model uses the gravitational perturbation the planet exhibits onto the stellar atmosphere \citep{Cuntz2000}, and the third model calculates the angular momentum transfer rate in the star-planet system \citep{Penev2012}.

\subsubsection{Tidal dissipation timescales for Spin-Orbit alignment}
\label{timescale}
In this model, the fundamental approach is to consider time scales for spin-orbit alignment in stellar binaries. It is observed that cool stars with close-in giant planets often have a low obliquity of the planetary orbital plane and the stellar spin, i.e.\ those planets tend to orbit in the stellar equatorial plane \citep{Albrecht2012}, which is interpreted as a consequence of tidal star-planet interaction. In contrast, hot host stars with a radiative outer layer display a wide range of obliquities. It is hypothesized that cool stars ultimately come into alignment with the orbits because they have higher rates of tidal dissipation due to the deeper convective zone \citep{Winn2010}. Hot stars, on the other hand, lack thick convective envelopes and have much longer tidal time scales.

Other studies, however, have shown that alignment between the orbital and stellar equatorial plane, which occurs predominantly in cool-star systems, is not necessarily linked to the tidal interaction between the star and its planet. \cite{Mazeh2015} have shown, in addition to the hot-cool dichotomy, that the distribution of the amplitude of photometric rotational modulation of Kepler Objects of Interest (KOIs), which is used as a proxy for the obliquity in star-planet systems, is similar for both short period and long period planets, demonstrating the possibility that tidal interaction may not be responsible for the alignment process. Further, \cite{Dai2017} investigated WASP-107b and found that both spin-orbit alignment and anti-alignment can be ruled out, although the planet is in a relatively tight orbit with a mass of $0.12 M_{Jup}$. Additionally, the authors found that, for a sample of stars with measured obliquities, hot stars are more likely to be misaligned and that cool stars have low obliquity when their planets are close-in, but seems independent of the mass ratio in the system. Following these findings, the authors conclude that scenarios involving tidal
realignment are questionable, but, additionally state that there is not any proposed mechanism that can explain the observed obliquities.

Going back to the work of \cite{Albrecht2012}, the dependence of obliquity of cool stars on the mass ratio was not evident. However, by invoking the tidal timescale formalism and comparing it to the stellar obliquity, a correlation supporting star-planet tidal interaction was found. Therefore, despite the concern regarding the origin of the alignment being in tidal interaction, we find that by comparing tidal dissipation timescale to the planet host activity excess, we present a methodology that tries to validate or discard the idea of tidal SPI independently of the stellar obliquity.

The formulae for tidal dissipation timescales are adopted from the spin-orbit synchronization timescales in binary star systems \citep{Zahn1977}. The following relationships between the star-planet system parameters and the convective (CE) and radiative\footnote{Radiative in the sense that stars hotter than 6250 K have an outer radiative envelope.} (RA) timescales for alignment are obtained:

\begin{equation}
\label{eq:timescales1}
\frac{1}{\tau_{CE}} = \frac{1}{10 \times 10^9 yr}q^2 \left( \frac{a/R_*}{40}  \right)^{-6}
\end{equation}
and,
\begin{equation}
\label{eq:timescales2}
\frac{1}{\tau_{RA}} = \frac{1}{0.25 \times 5 \times 10^9 yr}q^2(1+q)^{5/6} \left( \frac{a/R_*}{6} \right)^{-17/2},
\end{equation}
where \textit{q} is the planet-to-star mass ratio, $R_*$ is the stellar radius, and \textit{a} is the semi-major axis of the planetary orbit.

\subsubsection{Gravitational perturbation model}
\label{grav_pert}
The idea behind this model is that Hot Jupiters gravitationally influence the outermost atmospheric layers of their stars by raising tidal bulges at the subplanetary point \citep{Cuntz2000}. This may affect both the motion in the stellar convective layer and the flow fields in the outer atmosphere via the tidal bulges if the orbital and rotational periods are not equal. It is expected that this could lead to increased stellar activity because, in the case of different periods of the planet's orbit and the stellar spin, the stellar tidal bulges rise and subside quickly in the stellar rest frame, potentially increasing turbulent motions in the outer convection layer.

The parameter used to describe the SPI in this tidal interaction model by \cite{Cuntz2000} is the gravitational perturbation parameter $\Delta g_* / g_*$, which is defined as

\begin{equation}
\frac{\Delta g_*}{g_*} = \frac{M_p}{M_*}\frac{2R^3_*}{(a-R_*)^3}.
\end{equation}

Here, $M_*$ and $R_*$ are the stellar mass and radius, respectively, $M_p$ is the planetary mass, and \textit{a} is the semi-major axis of the planetary orbit. The gravitational perturbation parameter has the strongest dependence on the distance between the stellar surface and the planet $a-R_*$.


\subsubsection{Angular momentum transfer rate in Star-Planet System}
\label{torque}
The work of \cite{Penev2012} examines the efficiency of tidal dissipation by exoplanets in close-in circular orbits around stars with a surface convective zone. It does so by modeling the rotational evolution of a star that tidally interacts with a close-in planet. It also takes into account the coupling of the stellar convective envelope to the core and angular momentum loss due to the stellar wind. The Eqs.~\ref{eq:ang_mom1} and \ref{eq:ang_mom2} describe the angular momentum transfer rate and the evolution of the semi-major axis given the current value of planetary orbital period and stellar rotation period:

\begin{equation}
\label{eq:ang_mom1}
\left( \frac{dL_{conv}}{dt} \right)_{tide} = -\frac{1}{2}m_p M_* \sqrt{\frac{G}{a(M_*+m_p)}}\frac{da}{dt}
\end{equation}

\begin{equation}
\label{eq:ang_mom2}
\frac{da}{dt} = \text{sign}(\omega_{conv}-\omega_{orb})\frac{9}{2}\sqrt{\frac{G}{aM_*}} \left( \frac{R_*}{a} \right)^5 \frac{m_p}{Q_*},
\end{equation}

Here, $m_p$ and $M_*$ are the planet and stellar masses, respectively, \textit{G} is the gravitational constant, \textit{a} is the semi-major axis of the planet orbit, $\omega_{conv}$ and $\omega_{orb}$ are the angular frequency of the stellar convective zone and orbital angular frequency, respectively. Here, $\omega_{conv}$ is approximated by the observed stellar surface rotation. \cite{Penev2012} estimated that the tidal quality factor $Q_*$ has a value between $10^5$ and $10^7$. The lower limit implies a higher tidal dissipation efficiency, while the upper limit implies weaker efficiency. By choosing the efficiency to be $Q_* = 10^7$, we adopt a more conservative scenario which assumes that the weakest possible tides are acting.

The factor $\text{sign}(\omega_{conv}-\omega_{orb})$ takes the value $1$ when the stellar convective zone is spinning faster than the planet goes about its orbit and the value $-1$ in the opposite case. Therefore, this factor sets the direction of the angular momentum transfer.

To calculate $\omega_{conv}$, we used the stellar rotation periods from the literature (see table \ref{tab:stellar_planet_params}), assuming rigid body rotation: $\omega_{conv} = 2\pi/P_{rot}$. We took care to only consider stars for which the rotation period was measured directly from rotational modulation of the broad-band or chromospheric line emission, or from the rotational broadening of spectral lines. We did not use rotation periods that were derived from single-epoch activity measurements assuming a gyrochronology relationship, as such relationships may not be valid in the presence of star-planet interactions.

One constraint of the model is that it is only valid for low stellar masses. For masses larger than approximately $1.2 M_{Sun}$ the surface convective zone becomes negligible in mass. However, none of our sample stars has an estimated mass larger than $1.2 M_{Sun}$.


\begin{table*}
\begin{tabular}{lccccccc}
\hline
\hline
SP pair & $P_{rot}$[days] & $R_{*}$[$R_{\odot}$]  & $M_{*}$[$M_{\odot}$] & $P_{orb}$[days] & $M_{pl}$[$M_{Jup}$] & \textit{a}[AU] & $T_{eff}$[K] \\
\hline
16 Cyg Bb & $23.2 \pm 11.5 ^a$ & $1.13 \pm 0.08$ & $1.03 \pm 0.15$ & $798 \pm 1$ & $1.78 \pm 0.08$ & $1.66 \pm 0.03$ & $5747 \pm 143$\\[0.2cm]

55 Cnc Ae & $38.8 \pm 0.05 ^b$ & $0.96 \pm 0.08$ & $0.9 \pm 0.1$ & $0.736547 \pm 10^{-6}$ & $0.025 \pm 0.001$ & $0.0154 \pm 10^{-4}$ & $5250 \pm 171$\\[0.2cm]

CoRoT-2 Ab & $4.52 \pm 0.02 ^c$ & $0.94 \pm 0.12$ & $0.97 \pm 0.14$ & $1.742994 \pm 10^{-6}$ & $3.47 \pm 0.22$ & $0.02798 \pm 0.0008$ & $5529 \pm 121$\\[0.2cm]

GJ-15 Ab & $44.8^d$ & $0.41 \pm 0.01$ & $0.4 \pm 0.02$ & $11.441 \pm 0.002$ & $0.0095 \pm 0.0014$ & $0.072 \pm 0.004$ & $3607 \pm 68$\\[0.2cm]

HAT-P-20 Ab & $14.48 \pm 0.02 ^e$ & $0.68 \pm 0.07$ & $0.73 \pm 0.1$ & $2.875317 \pm 10^{-6}$ & $7.2 \pm 0.2$ & $0.0361 \pm 0.0005$ & $4604\pm 129$\\[0.2cm]

HD 46375 Ab & / & $0.92 \pm 0.05$ & $0.9 \pm 0.13 $ & $3.0236 \pm 10^{-4}$ & $0.23 \pm 0.02$ & $0.0398 \pm 0.0023$ & $5092 \pm 149$\\[0.2cm]

HD 75289 Ab & $16.0^f$ & $1.3 \pm 0.07$ & $1.3 \pm 0.1$ & $3.5093 \pm 10^{-4}$ & $0.49 \pm 0.03$ & 0.05 & $6044 \pm 156$\\[0.2cm]

HD 109749 Ab & / & $1.3 \pm 0.07$ & $1.06 \pm 0.17$ & $5.2399 \pm 0.0001$ & $0.27 \pm 0.05$ & $0.062 \pm 0.004$ & $5868 \pm 162 $\\[0.2cm]

HD 178911 Bb & / & $1.02 \pm 0.06$ & $0.98 \pm 0.18$ & $71.48 \pm 0.02$ & $8.03 \pm 2.51$ & $0.34 \pm 0.01$ & $5564 \pm 163$\\[0.2cm]

HD 188015 Ab & / & $1.084 \pm 0.075$ & $1.02 \pm 0.15$ & $461.2 \pm 1.7$ & $1.5 \pm 0.13$ & $1.203 \pm 0.07$ & $5727 \pm 139$ \\[0.2cm]

HD 189733 Ab & $11.95 \pm 0.01 ^g$ & $0.78 \pm 0.06$ & $0.84 \pm 0.11$ & $2.2185757 \pm 10^{-7}$ & $1.13 \pm 0.08$ & $0.0311 \pm 0.0005$ & $5023 \pm 120$\\[0.2cm]

HD 190360 Ac & / &$ 1.17 \pm 0.07$ & $0.98 \pm 0.07$ & $17.111 \pm 0.005$ & $0.06 \pm 0.01$ & $0.130 \pm 0.008$ & $5549 \pm 123$\\[0.2cm]

$\upsilon$ And Ab & $12.0^h$ & $1.64 \pm 0.11$ & $1.15 \pm 0.16$ & $4.61703 \pm 10{-5}$ & $0.688 \pm 0.004$ & $0.0592217 \pm 10^{-7}$ & $6183 \pm 35$\\[0.2cm]

WASP-8 Ab & $16.4 \pm 1.0 ^i$ & $0.997 \pm 0.06$ & $0.99 \pm 0.16$ & $8.15872 \pm 10^{-5}$ & $2.54 \pm 0.33$ & $0.0801 \pm 0.0015$ & $5589 \pm 174$\\[0.2cm]

WASP-77 Ab & $15.4 \pm 0.4 ^j$ & $0.93 \pm 0.05$ & $0.99 \pm 0.16$ & $1.36003 \pm 10^{-5}$ & $2.29 \pm 0.33$ & $0.0241 \pm 0.0004$ & $5605 \pm 115$\\[0.2cm]

XO-2 Sb & $30.0 \pm 4.0 ^k$ & $0.96 \pm 0.05$ & $0.98 \pm 0.12$ & $18.16 \pm 0.03$ & $0.26 \pm 0.04$ & 0.13 & $5547 \pm 118$\\[0.2cm]

XO-2 Nb & $41.6 \pm 1.1 ^l$ & $1.09 \pm 0.09$ & $0.9 \pm 0.13$  & $2.6158592 \pm 10^{-7}$ & $0.595 \pm 0.022$ & $0.0367 \pm 0.0006$& $5267 \pm 190$\\

\hline
\end{tabular}
\caption{Given are the stellar and planetary parameters used for the star-planet tidal interaction models. The stellar parameters ($M_*$, $R_*$, and $T_{eff}$) given here are taken from the TESS Input Catalog \citep{TICv8}, while rotation periods were researched individualy: $^a$: \protect\citep{Davies2015}; $^b$: \protect\citep{Bourrier2018}; $^c$: \protect\citep{Lanza2009corot}; $^d$: \protect\citep{Howard2014}; $^e$: \protect\citep{Esposito2017}; $^f$: \protect\citep{Daz2018}; $^f$:\protect\citep{Udry2000}; $^g$: \protect\citep{Henry2007}; $^h$: \protect\citep{Butler1997}; $^i$:\protect\citep{Salz2015}; $^j$: \protect\citep{Maxted2013}; $^k,^l$: \protect\citep{Damasso2015}. For the planetary parametars, we used the publications  referred in the \href{https://exoplanetarchive.ipac.caltech.edu/}{NASA Exoplanet Archive} \citep{Akeson2013}: {\bf 16 Cyg b}: \protect\citep{Stassun2017}, {\bf 55 Cnc e}: \protect\citep{Bourrier2018}, {\bf CoRoT-2 b}: \protect\citep{Gillon2010}, {\bf GJ-15 b}: \protect\citep{Pinamonti2018}, {\bf HAT-P-20 b}: \protect\citep{Bakos2011}, {\bf HD 46375 b}: \protect\citep{Butler2006}, {\bf HD 75289 b}: \protect\citep{Stassun2017}, {\bf HD 109749 b}: \protect\citep{Ment2018}, {\bf HD 178911 b}: \protect\citep{Stassun2017}, {\bf HD 188015 b}: \protect\citep{Butler2006}, {\bf HD 189733 b}: \protect\citep{Stassun2017, Bonomo2017}, {\bf HD 190360 c}: \protect\citep{Wright2009}, {\bf $\upsilon$ And b}: \protect\citep{Curiel2011}, {\bf WASP-8 b}: \protect\citep{Stassun2017, Bonomo2017}, {\bf WASP-77 b}: \protect\citep{Stassun2017, Bonomo2017}, {\bf XO-2 Sb}: \protect\citep{Stassun2017}, {\bf XO-2 Nb}: \protect\citep{Bonomo2017}. If the uncertainty of a stellar or planetary parameter was given as an asymmetrical interval in the corresponding publication, we used the greater value for further analysis.}
\label{tab:stellar_planet_params}
\end{table*}

\subsubsection{Model results}

We calculated the strength of the tidal star-planet interaction according to the three models for the planet-hosting stars in our sample and then compared them to the observed activity level difference to their same-age stellar companions.
Systems that have both stellar components undetected were discarded from this analysis. For these systems, we were not able to estimate an upper or lower limit to the activity level difference. For systems where we have at least one detection, we were able to give either a lower or an upper limit to the activity level difference, depending on which star was detected.  

Our activity level difference parameter is calculated as follows: we compare the X-ray luminosity of each individual star to the X-ray luminosity function for the respective spectral type and calculate into which percentile of the distribution the star falls. Then we calculate the difference in percentiles of two stars in a system with respect to their applicable X-ray luminosity functions. If a planet-hosting star is much more active than its companion star with respect to their spectral types, then this will yield a positive percentile difference, if it is much less active, a negative one. If the percentile difference is close to zero, both stars have similar activity levels for their respective spectral types.

\begin{figure*}
  \includegraphics[width=0.85\linewidth]{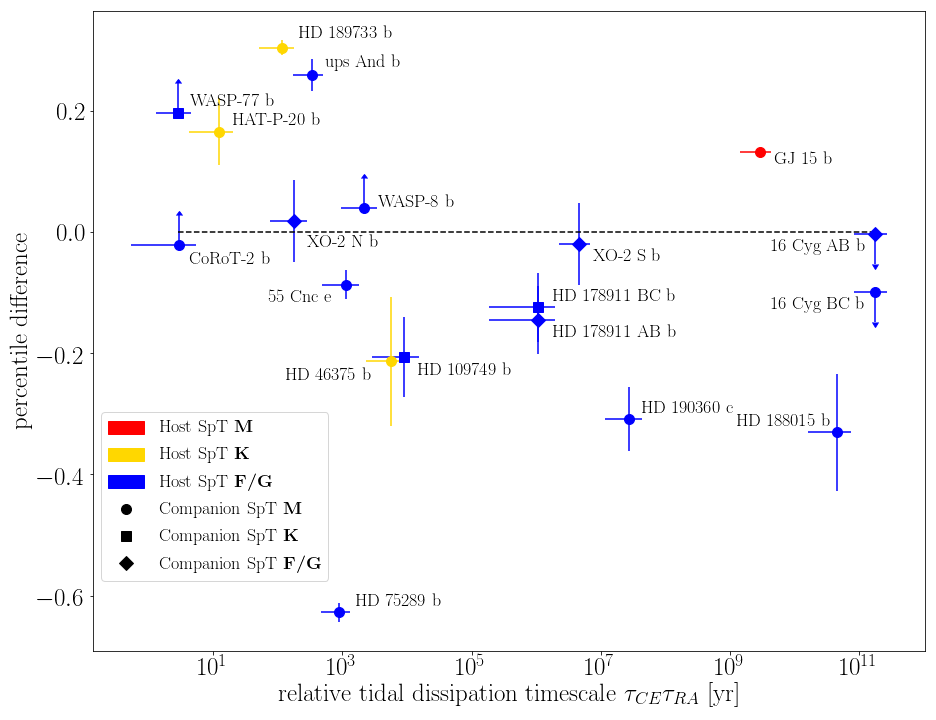}
  \caption{Shown is the percentile difference as a function of the tidal dissipation timescale for each star-planet system from our final sample. Here, the star-planet interaction model described in section \ref{timescale} was used. Color-coded is the planet host spectral type, while the shape of the marker indicates the spectral type of the stellar companion. The tidal timescales are given relative to the solar age.}
  \label{fig:timescales}
\end{figure*}

\begin{figure}
  \includegraphics[width=\linewidth]{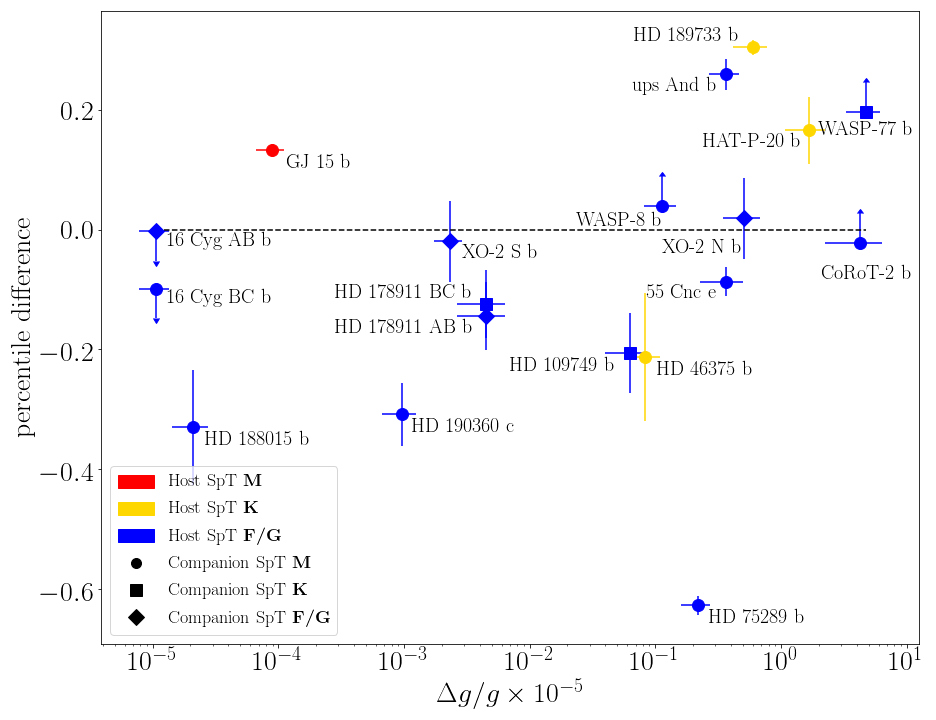}
  \caption{This figure shows the percentile difference as a function of the gravitational perturbation parameter described in section \ref{grav_pert}. As in Figs. \ref{fig:timescales}, color-coded is the planet host spectral type, while the shape of the marker indicates the spectral type of the stellar companion.}
  \label{fig:grav_pert}
\end{figure}

\begin{figure}
  \includegraphics[width=\linewidth]{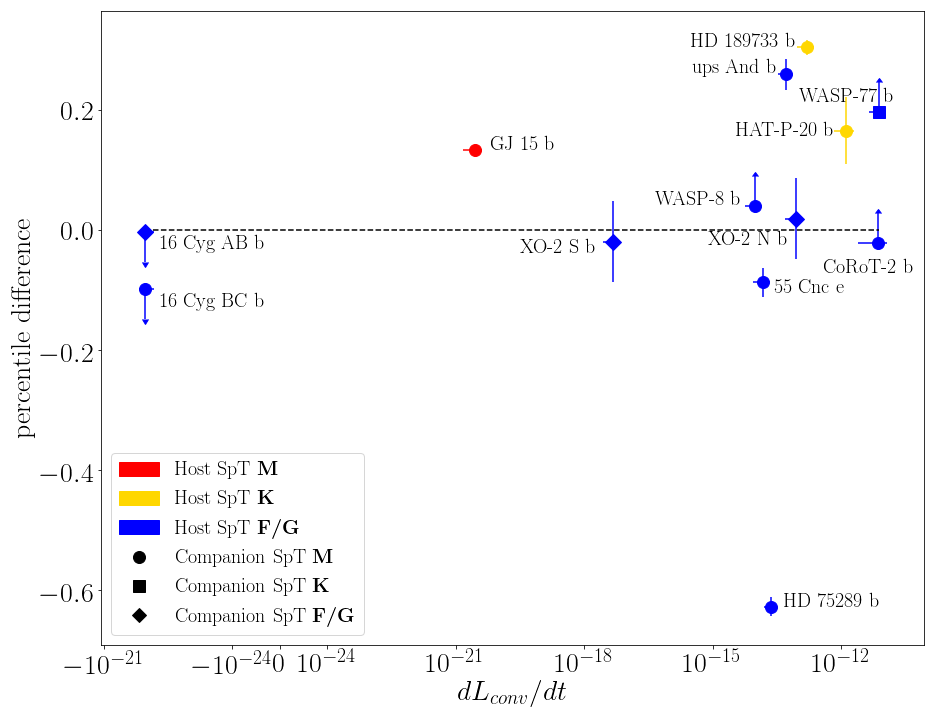}
  \caption{Here, the percentile difference as a function of the torque that is transfered between the planet's orbit and stellar spin is given for each system from our final sample. The tidal interaction model described in section \ref{torque} was used to calculate the tidal interaction parameter used. As in previous result figures, color-coded is the planet host spectral type, while the shape of the marker indicates the spectral type of the stellar companion.}
  \label{fig:ang_mom}
\end{figure}

Figures \ref{fig:timescales}, \ref{fig:grav_pert}, and \ref{fig:ang_mom} show the activity level difference as a function of the star-planet tidal interaction parameters of the three models described above. In the result figures, we have color-coded the information on the spectral type of the planet host, whereas the symbol shape accounts for the spectral type of the companion. The percentile difference uncertainty is estimated by applying the error propagation function on the X-ray luminosity uncertainty of both stellar components. This is done for systems where both components are detected.

To assess the relationship between the activity level and the interaction strength, we employ the statistical Spearman's rank analysis. In statistics, Spearman's rank correlation coefficient $\rho$ is a nonparametric measure of rank correlation. It assesses how well the relationship between two variables can be described using a monotonic function of arbitrary form. For this analysis, we used systems where both components are detected, and those where the host star is detected and the stellar companion is undetected, i.e.\ where we have a lower limit to the activity level difference. We discarded the \textit{16 Cyg} system, where the planet host was undetected, as we were only able to estimate the upper limit of the percentile difference for this system.

The correlation coefficients for the three SPI models are given in Table \ref{table:sp_parameters}.
The corresponding p-value gives the probability that the observed value of $\rho$ can be obtained by statistical fluctuations. We find that the first model, where short time scales mean strong tidal interactions, shows strong anticorrelation as expected, the second model shows a highly significant correlation, and the third model shows a mild correlation. We point out that the sample that could be used for model three was smaller since it required the rotation period of the host star to be known.



In our sample were five known multiplanet systems: WASP-8 A, 55 Cnc A, HD 190360 A, $\upsilon$ And A, and XO-2 S. We calculated the SPI strength for each star-planet pair; for the WASP-8 A, $\upsilon$ And A, and XO-2 S planet systems, the b planet was the one yielding a stronger influence on the planet host in all three models, while in the system HD 190360 A this was true for the c component. For the 55 Cnc A system, the planet e yielded the strongest influence on the host star.

\begin{table*}
\begin{tabular}{lcccccccccccccccccccc}
\hline
\hline
system & percentile difference & $\Delta g/g$ & tidal timescale$\left[\frac{yr}{5\times10^9} \right]$ & $dL_{conv}/dt \left[ M_{\odot}\left(\frac{km}{s}\right)^2 \right]$ \\
\hline
16 Cyg AB & $\leq$ -0.0023 & \multirow{2}{*}{$(1.057 \pm 0.283)\times 10^{-10}$} & \multirow{2}{*}{$(1.783 \pm 0.952)\times 10^{11}$} & \multirow{2}{*}{$-(1.101 \pm 0.420)\times 10^{-22}$}   \\
16 Cyg BC & $\leq$ -0.0984 &  &  &  \\[0.2cm]
55 Cnc & $-0.0867 \pm 0.0239$ & $(3.647\pm1.379) \times 10^{-6}$ & $1149.778 \pm 669.645$ & $(1.529 \pm 0.631)\times 10^{-14}$ \\[0.2cm]
CoRoT-2 & $\geq$ -0.0217 & $(4.294 \pm 2.038)\times 10^{-5}$ & $2.928 \pm 2.398$ & $(7.187 \pm 4.665)\times 10^{-12} $ \\[0.2cm]
GJ-15 & $0.1331 \pm 0.0042$ &  $(8.928 \pm 2.245)\times 10^{-10}$ & $(2.874 \pm 1.423)\times 10^9 $& $(2.841 \pm 1.349)\times 10^{-21} $\\[0.2cm]
HAT-P-20 & $0.166 \pm 0.0558$ & $(1.661 \pm 0.585)\times 10^{-5} $& $12.251 \pm 7.994$ & $(1.336 \pm 0.663)\times 10^{-12} $  \\[0.2cm]
HD 46375 & $-0.2133 \pm 0.1069$ & $(8.316 \pm 2.618) \times 10^{-7}$ & $5579.114 \pm 3247.165$ & /  \\[0.2cm]
HD 75289 & $-0.6278 \pm 0.0159$ &  $(2.173 \pm 0.560)\times 10^{-6} $& $895.898 \pm 433.828$ & $(2.254 \pm 0.667)\times 10^{-14} $  \\[0.2cm]
HD 109749 & $-0.2063 \pm 0.0664$ & $(6.301 \pm 2.293)\times 10^{-7} $& $9152.284 \pm 6278.918$ & / \\[0.2cm]
HD 178911 AB & $-0.1443 \pm 0.0563$ & \multirow{2}{*}{$(4.484 \pm 1.855)\times 10^{-8}$} & \multirow{2}{*}{$(1.057 \pm  0.872)\times 10^6$} & \multirow{2}{*}{/}   \\
HD 178911 BC & $-0.1243 \pm 0.0563$ &  &  &   \\[0.2cm]
HD 188015 & $-0.3305\pm 0.0967$ & $(2.092 \pm 0.677)\times 10^{-10} $& $(4.575 \pm 2.951)\times 10^{10} $& / \\[0.2cm]
HD 189733 & $0.3049 \pm 0.0126$ & $(5.977 \pm 1.841)\times10^{-6}$ & $115.361 \pm 64.808$ & $(1.640 \pm 0.699)\times 10^{-13}$  \\[0.2cm]
HD 190360 & $-0.3084 \pm 0.0530$ & $(9.598 \pm 2.860)\times 10^{-9}$ & $(2.737\pm1.579)\times 10^7 $& / \\[0.2cm]
$\upsilon$ And & $0.2597 \pm 0.0258$ & $(3.656 \pm 0.961)\times 10^{-6}$ & $333.397 \pm 159.447$ & $(5.073\pm 1.648)\times 10^{-14} $  \\[0.2cm]
WASP-8  & $\geq$ 0.0397 & $(1.136 \pm 0.328)\times 10^{-6}$ & $2164.122 \pm 1214.617$ & $(9.496 \pm 3.922)\times 10^{-15} $  \\[0.2cm]
WASP-77 & $\geq$ 0.1964 & $(4.729 \pm 1.430)\times 10^{-5} $& $2.886 \pm 1.599$ & $(7.602 \pm 3.121)\times 10^{-12} $  \\[0.2cm]
XO-2 S & $-0.0193 \pm 0.0677$ & $(2.307\pm 0.585)\times 10^{-8} $& $(4.527 \pm 2.264)\times 10^6$ & $(4.582 \pm 1.846)\times 10^{-18} $ \\[0.2cm]
XO-2 N & $0.0193 \pm 0.0677$ & $(5.125 \pm 0.165)\times 10^{-6} $& $180.842 \pm 104.043$ & $(8.676 \pm 3.652)\times 10^{-14}$  \\[0.2cm]
\hline
\end{tabular}
\caption{Given are the tidal interaction strength values and their uncertainties for all three star-planet interaction models used. Regarding the uncertainty given for the angular momentum transfer rate, the rotational period uncertainty was not taken into account since it was never high enough to change the sign of the torque. For the triple systems 16 Cyg and HD 178911, we were able to estimate the X-ray luminosity of all components and, therefore, calculated the activity level difference for each pair.}
\label{tab:percentiles_and_models}
\end{table*}

\begin{table}
\begin{tabular}{cccc}
\hline
\hline
& \multicolumn{3}{c}{SPI models} \\
 & tidal timescale & $\Delta g/g$ & $dL_{conv}/dt$ \\
\hline
Spearman's $\rho$ & -0.5382 & 0.5559 & 0.3091 \\

p-value & 0.0315 & 0.0254 & 0.3550 \\
\hline
\end{tabular}
\caption{Given are the Spearman's rank correlation coefficients for all three star-planet interaction models we used, together with the p-value for each given set of data points.}
\label{table:sp_parameters}
\end{table}

\section{Discussion}
\label{discussion}
\subsection{Activity bias in planet detections}

In general, there exists a detection bias toward close-in giant planets (Hot Jupiters) around more active stars. The radial velocity signal of smaller and/or distant planets is more difficult to detect around active stars because of the low signal-to-noise ratio \citep{Saar1997,Hatzes2002,Desort2007,Lagrange2010}. Therefore, if we compare all activity measurements of the full planet-hosting star sample to one another, we would not be properly accounting for the detection bias. However, if each host star is compared to some age or activity expectation independent of the parameters of the host itself, the detection bias can be overcome. Here, our wide binary stellar systems fulfill that purpose. If a Hot-Jupiter-hosting star has a high activity level, using the stellar companion, we can determine if the activity is due to the young age -- the companion then also has a high activity level for the given spectral type (see e.g. \cite{Johnstone2020} and references therein) -- or due to star-planet interaction -- the companion then has a significantly lower activity level. With this approach, each planet host has its own activity proxy and the detection bias is properly accounted for.

Also, regarding the age of a system, detecting planets around older stars is somewhat easier, especially with the radial velocity method since older stars tend to be slower rotators with lower magnetic activity. However, since we analyse stellar X-ray activity of the host star {\it relative} to the other star in the wide binary system, it is not of importance what the true intrinsic age of those systems is since we only compare one star to one other star with the same age.

\subsection{Interpretation of the observed activity difference}

To avoid the activity bias, we have chosen wide binary systems, as discussed previously. Our results show a positive correlation between the magnetic activity level of the planet-hosting star relative to its stellar companion and the star-planet tidal interaction strength in the second and third model. For the first model, we expect an anticorrelation between the tidal timescale and the activity difference, as short timescales indicate strong tidal interaction, and this is indeed what we find. 


For illustrative purposes, we can  now compare our findings to what we would expect to see if the result was purely driven by any remaining effect of the planet detection bias with respect to the activity. Under a planet detection bias, we expect that Hot Jupiters (i.e.\ tidally strongly interacting planets) are found around both active and inactive host stars, and small planets in wide orbits (i.e.\ tidally weakly interacting planets) are found mainly around inactive stars. In our Figure \ref{fig:timescales} this would manifest itself as a population of systems that fill the plot in a triangular shape to the lower left, i.e.\ the upper right of the plot would be devoid of systems. Conversely, for Figs.~\ref{fig:grav_pert} and \ref{fig:ang_mom} the sample should fill the lower right part of the plot in a triangular shape, meaning that only small, far-out planets around active stars are underrepresented.

However, we see that the strongly interacting planets, i.e.\ the Hot Jupiters, are actually not found around stars that have any arbitrary activity level in relation to their stellar companion but are primarily found around host stars that tend to be as active or more active than their companion stars. 
In a more quantitative manner, out of the eight Hot-Jupiter SP systems in our final sample, we find that 75\% of them orbit host stars that have an excess in coronal activity, while the remaining 25\% are found around host stars with no significant activity excess.\footnote{Here, we define Hot Jupiters to be planets with an orbital period less than 10 days and a mass higher than $0.25 M_{Jup}$ \citep{Dawson2018}. The Hot Jupiters in our sample are then: CoRoT-2 A, HAT-P-20 A, HD 75289 A, HD 189733 A, $\upsilon$ Andromedae A, WASP-8 A, WASP-77 A, and XO-2 N; the planet orbiting HD 109749 A may have a mass lower than the given hot-Jupiter limit, considering its mass uncertainty.}
Therefore, we argue that the observed relation is indeed a signpost of star-planet interaction, and not due to activity biases in exoplanet detections.

One surprising feature of our sample is the fact that host stars of small, far-out planets seem to be on average \textit{less} active than their companion stars, even after we have corrected for spectral type differences (i.e.\ the lower right corner in Fig.~\ref{fig:timescales}). This trend can be explained by the different main-sequence lifetimes of F to M stars. Several of those low-tidal interaction host stars are of spectral type F/G, while the companion star is of spectral type M (codified by symbol colours and shapes in Figs.~\ref{fig:timescales}, \ref{fig:grav_pert}, and \ref{fig:ang_mom}). For old and inactive F/G stars, where those low-tidal interaction planets are more easily detected, the system age may already be close to the main-sequence lifetime of the primary. However, the secondary M dwarf will not yet be at the end of its main-sequence lifetime. In terms of those two stars' position in the X-ray luminosity functions for their spectral type, the F/G star will be at the faintest end of the luminosity function, while the M dwarf will not have moved fully to the faint end of the M dwarf luminosity function yet. This will cause a net negative activity level difference.

Two of the most inactive systems are HD 190360 and HD 188015 (HD 75289 is discussed separately).
The primary components of the two systems are G-dwarfs, while the secondaries are M-dwarfs. Also, both systems seem to be older than the solar age.
\cite{Takeda2007} estimated the age of HD 190360 and HD 188015 to be 13.4 and 6.2 Gyr, respectively.

We point out that there are also a number of F/G planet host stars with M dwarf secondaries in the high-tidal interaction part of the sample (i.e.\ at the upper left corner in Fig.~\ref{fig:timescales}). Yet there, the primaries are found to be more active than the secondaries. Therefore we are confident that it is not spectral type differences that are driving the trends in our sample.

We also point out that our sample does not allow us to test for different behaviours among stars with a convective versus a radiative envelope, as was done by \cite{Albrecht2012} and \cite{Winn2010}. Our sample does not contain host stars with masses above $1.2M_\odot$, therefore the different tidal efficiencies of model 1 cannot be tested observationally here.

\subsubsection{The low-activity outlier HD 75289}
\label{HD75289}
One strong outlier from the trend established by the other systems is the system HD~75289, which is an F9 star hosting a Hot Jupiter and having an M dwarf companion. The host star is very inactive compared to the lower-mass stellar companion, and we did not find evidence of a flare happening on the M dwarf during the X-ray observations. In fact, the X-ray luminosity of the primary is within the lowest 1\% for F/G-dwarfs, while the M-dwarf companion has an average field star luminosity value.

\cite{Takeda2007} estimated the age of the system to be 3.28 Gyrs, making this system younger than the Sun. The rotational period of the primary has been measured from the chromospheric activity level and radial velocity to be ca.\ 16 days \citep{Udry2000}, see Table \ref{tab:stellar_planet_params}. When comparing the rotation and X-ray luminosity of the primary to the sample of field stars studied for the rotation-activity relationship by \cite{Pizzolato2003}, we find that this star's activity is extremely low for its rotational state (see Fig.~\ref{fig:pizzolato}). We speculate that this star may be undergoing a low-activity phase in a magnetic cycle, or possibly be in a Maunder minimum state \citep{Eddy1976}.


\subsection{Tidal or magnetic star-planet interaction?}

\begin{figure}
  \includegraphics[width=\linewidth]{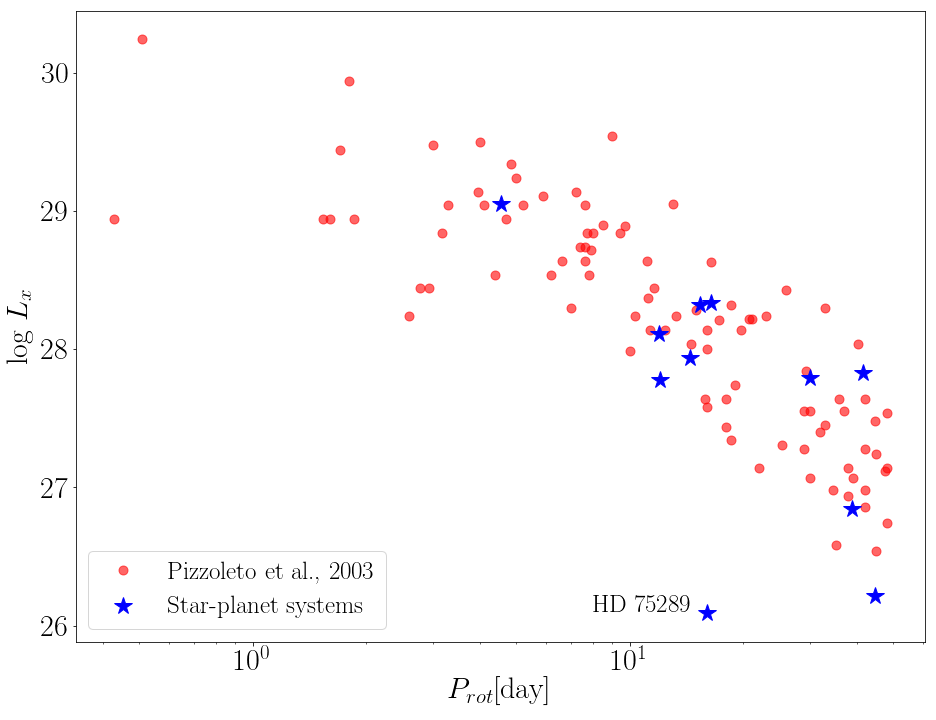}
   \caption{Shown is the field-star sample from \protect\cite{Pizzolato2003} together with planet-hosting stars from our analysis that had their rotational period estimated in the literature. Since the field-star sample was observed with the ROSAT space telescope, the conversion between the two relevant energy bands given by the Eq. \ref{ROSAT_conversion} was also applied here.}
  \label{fig:pizzolato}
\end{figure}
Assuming that massive planets, as the jovian planets in the Solar system, have magnetic fields, being close to their host stars may induce magnetic interaction in addition to tidal interaction. It is therefore imaginable that the activity discrepancy we observe in our binary sample is, to some part, induced by this interaction. The rotation-activity relationship of field stars has been explored in X-rays by many studies \citep{Pallavicini1981,Pizzolato2003,Wright2011}. We have chosen to show the field stars sample from \cite{Pizzolato2003} in Fig. \ref{fig:pizzolato}.

We show the planet-hosting stars from our sample, for which rotation periods are known and the X-ray luminosity has been measured here, in addition to the Pizzolato sample. With the exception of HD 75289 which was discussed above, the host stars do not deviate significantly from the relationship seen in regular field stars. This means that the rotation and activity are in lockstep for our planet-hosting stars. If short-term magnetic star-planet interaction was a strong effect in our sample, we would expect the stars to have elevated activity while their rotation remains unchanged. As this is not the case, we argue that short-term magnetic star-planet interaction does not play a major role in our sample.

However, also the existence of long-term magnetic star-planet interaction has been proposed \citep{Cohen2010,Strugarek2014}. In such a scenario, the magnetic interaction between the star and the planet can modify the stellar wind properties. In particular, the angular momentum loss due to the stellar wind might be decreased and the star retains a higher rotation rate and therefore a higher activity level. This scenario is not excluded by our considerations here, since then both activity and rotation could be changed over the stellar lifetime. One argument against this scenario is however that the systems in which the strongest magnetic interaction is expected, namely the Hot Jupiter systems, may have weak magnetic fields of the exoplanets. This was posited by \cite{Griessmeier2004} due to the tidal locking of exoplanet spin to their orbital period. In the case of a weak or non-existent planet magnetosphere, the long-term magnetic star-planet interaction is less effective than the tidal star-planet interaction we assume here.

\subsection{Possible Caveats}

One phenomenon that we were not able to take explicitly into account is potential stellar magnetic activity cycles. It is observed that stars, when going through activity cycles, can also exhibit variation in their X-ray luminosity \citep{Robrade2012}. At first glance, this may be viewed as an issue, but it is statistically unlikely that all the stars showing a higher activity level are in their maximum of the activity cycle while their companions experience an activity minimum. The same can be assumed in the opposite case, where the host stars of small and/or further out planets would have a low activity level due to an activity minimum.

One other magnetic phenomenon that may increase the X-ray luminosity of a star is flares.
For systems observed with the Chandra space observatory, we were not able to properly investigate the X-ray light curves for flaring events due to the low number of counts over time\footnote{Systems HD~189733 and CoRoT-2 were analysed in more detail by \cite{Poppenhaeger2013} and \cite{Schroter2011}, respectively, and no flaring events were detected}. Here, the possibility exists that these planet-hosting stars appear more active because a flare occurred during the observation time. However, it is again unlikely that in the higher activity level regime, all planet hosts experienced a flaring event, while in systems in the lower-activity level regime the planet host companions experienced flares that made them appear more active.

Therefore, we assume that the activity cycles and possible undetected flaring events add to the scatter we see in our activity-tidal interaction strength figures; however, they do not produce the correlation we see.

\section{Conclusions}
\label{conclusion}
As can be seen from the result figures and the Spearman's rank correlation factors, for all three star-planet interaction models there is a strong correlation between the SPI strength parameter and the activity level difference for a given star-planet system residing in a binary system. If we examine more closely emerging correlation, we see that in systems where a massive planet has a tight orbit, the planet-hosting star tends to be more active, i.e. have a hotter corona. If we assume that the activity level difference between the stellar components of a system is not correlated with the tidal interaction strength, we would have star-planet systems where an inactive host is orbited by a close-in massive planet. That would indicate that for a similar tidal interaction strength, we would have planet-hosting stars that range from very active (positive percentile difference) to very inactive (negative percentile difference) when compared to its stellar companion's activity level. This would indicate, that the planet does not influence the behavior of its host in any way. Since in our case the low activity-high interaction regime is scarcely populated, tidal star-planet interaction can be efficient in altering the stellar spin evolution and therefore altering its activity level.

\section*{Acknowledgements}

The authors thank Judy Chebly for discussions on optical data of some of the sample stars.
Parts of this work were supported by the German \emph{Leibniz-Gemeinschaft} under project number P67/2018.
The scientific results reported in this article are based in part on observations made by the Chandra X-ray Observatory and the X-ray Multi-Mirror Observatory XMM-Newton. This research made use of Astropy,\footnote{http://www.astropy.org} a community-developed core Python package for Astronomy \citep{Astropy2013, Astropy2018}. 
This work has made use of data from the European Space Agency (ESA) mission
{\it Gaia} (\url{https://www.cosmos.esa.int/gaia}), processed by the {\it Gaia}
Data Processing and Analysis Consortium (DPAC,
\url{https://www.cosmos.esa.int/web/gaia/dpac/consortium}). Funding for the DPAC
has been provided by national institutions, in particular the institutions
participating in the {\it Gaia} Multilateral Agreement.

\section*{Data availability}

The data in this article are available from the XMM-Newton Science Archive (\url{https://www.cosmos.esa.int/web/xmm-newton/xsa}) with the observation IDs as given in Table~\ref{table:1} and from the Chandra X-ray Center (\url{https://cxc.cfa.harvard.edu/}) with the observation IDs as given in the Table~\ref{table:3}. The data products generated from the raw data are available upon request from the author.




\bibliographystyle{mnras}
\bibliography{references} 




\appendix

\section{Notes on individual systems}
\label{appendA}

\textit{16 Cyg:} This is a hierarchical triple star system, where the AC components form a close binary and the B component is the planet host. There are two observations of this system with XMM-Newton (obsID: 0551021701 and 0823050101). The former observation has a short exposure time and yields only weak or no detection of the components. The latter observation yields two detections where the radiation hardness ratio of the planet host and the unresolved binary system was estimated, therefore we used only this observation for the primary estimate of the X-ray luminosity of the components given in Table \ref{tab:XMM_Newton_flux}. This system was also observed with Chandra's ACIS-I instrument. Here, the three components are resolved, but the planet host is undetected. We used the Chandra observation for the estimate of the percentile difference between the stellar components since the XMM observations yielded the AC pair as unresolved. The angular separation between the planet host and the binary system is $\rho=40.0\arcsec$, while the separation between the components of the binary is $\rho \approx 3.5 \arcsec$.

\textit{30 Ari (HD 16246):} The system is a hierarchical triple system (possible quadruple \citep{RobertsJr2015}). The planet host is the B component, which has the C component in close orbit. The angular separation between the components B and C is $\rho = 0.536\arcsec$, which makes them unresolved in the XMM-Newton observation. The C component is an M1-3 spectral type star with an orbital period of $P\le 95$ yr \citep{Kane2015}. The angular separation between the components A and B is $\rho = 37.9 \arcsec$ \citep{Mugrauer2019}. The system was detected in all three EPIC detectors as two bright X-ray sources.

\textit{55 Cnc (HD 75732):} Both components of the system were detected in all three EPIC detectors, but the hardness ratio of the radiation detected from the second component could not be estimated. A coronal temperature of $\log_{10}T=6.477$ was employed to estimate the X-ray luminosity of the secondary. The system was also observed with Chandra, but only the primary component was in the FoV. The angular separation between the two components is $\rho = 85.0\arcsec$.

\textit{83 Leo (HD 99491 and HD 99492):} Here, the secondary component is the planet host. Both components are detected in all three EPIC detectors: the primary is a bright and the secondary a faint X-ray source. The angular separation between the components is $\rho=28.0 \arcsec$.

\textit{AS 205}: The two components are unresolved in both XMM-Newton and Chandra observations, at an angular separation of $\rho = 1.3\arcsec$. In XMM-Newton observation, the system was detected in all three EPIC detectors, but the radiation hardness ratio could not be estimated. In Chandra's observation, the system was detected with an HR estimate.

\textit{CoRoT-2}: This system was observed with Chandra's ACIS-S instrument. The primary component was detected, while the secondary was not. Therefore, we estimated the $2-\sigma$ upper limit of the X-ray luminosity of the secondary. The angular separation between the components is $\approx 4 \arcsec$. Since this system was the topic of research in \cite{Schroter2011} where the observed X-ray spectrum was used to characterise the primary component, we used their published X-ray flux value of CoRoT-2 A for further analysis.

\textit{GJ-15 (HD 1326):} This system was observed with XMM-Newton and has two bright X-ray sources. The primary component has a flaring event with a duration of $\approx$ 3ks. We excluded this time interval from our calculation of the primary component's X-ray luminosity. The angular separation between the components is $\rho=34.4\arcsec$.

\textit{HAT-P-16:} This system is a hierarchical triplet, where the planet host has a close unresolved companion and the C component is at an angular separation of $\rho=23.3\arcsec$. There are two observations of this system with XMM-Newton (obsID: 0800733701 and 0800733101). The latter observation had a high background signal rendering the components undetected when combining the observations. Using only the former observation, we had a detection although without the hardness ratio of the radiation. Nevertheless, we were able to calculate the source flux, by assuming a coronal temperature of $\log_{10}T=6.477$.

\textit{HAT-P-20:} This system was observed with Chandra's ACIS-S instrument. Both components were detected, but the secondary had no HR estimate. The angular separation between the components is $\rho \approx 7 \arcsec$.

\textit{HAT-P-22 (HD 233731):} This system was observed by the Chandra Space Observatory. It was not the main target and it was positioned at an off-axis angle of $\approx 4 \arcmin$. Therefore, we used a somewhat larger source extraction region of $\approx 4.5 \arcsec$. Both companions are undetected, therefore the system was not used for further analysis. The angular separation between the components is $\approx 9 \arcsec$.

\textit{HATS-65:} This system was observed with Chandra's ACIS-I instrument. Its position was projected at an off-axis angle of $\approx 10 \arcmin$, which employed us to use a source extraction region with a radius of $10 \arcsec$ to collect most of the photon events coming from this system. The angular separation between the components is $\approx 5 \arcsec$, which renders this system unresolved. It was also undetected, therefore, we only give the $2-\sigma$ upper limit for the X-ray luminosity of this system as one source.

\textit{HD 27442:} This system has two observations with XMM-Newton (obsID: 0780510501 and 0551021401). The primary was detected in both observations, but only the obs no.\ 0780510501 yields a radiation hardness ratio. The secondary component was detected in the former observation with no HR estimate and in the latter observation, no detection was made. We combined the observations of both the primary and the secondary for a better estimate of the SNR. For the primary, we were able to calculate the HR, but we were not able to calculate a radiation hardness ratio for the secondary. We, therefore, assumed a coronal temperature of $\log_{10}T=6.477$ for this component. The latter observation also yields a large background signal. The angular separation between the components is $\rho=13\arcsec$ and the source extraction region of both components is $\approx 10 \arcsec$. We, therefore, could not avoid an overlap between the extraction regions of the two components. This system, however, consists of two evolved stars \citep{Butler2001,Mugrauer2007} and is, therefore, excluded from further analysis.

\textit{HD 46375:} The system was observed with XMM-Newton, in all three EPIC detectors, as unresolved as the components are at an angular separation of $\rho = 10.4 \arcsec$. It was also observed and detected by Chandra's ACIS-S instrument, where the system appeared as resolved.

\textit{HD 75289:} This system has two observations with XMM-Newton. The primary component was not detected in both observation runs therefore, we combined them to achieve a better SNR. With the combined observation, we detected the primary component, but no hardness ratio estimate was possible and a coronal temperature of $\log_{10}T=6.477$ was assumed. The secondary component was detected in both observation runs, but we were able to calculate the radiation hardness ratio for the detection in observation no.\ 0722030301. For the second component in observation no.\ 0304200501, we assumed the same HR as in the previous observation and calculated the X-ray luminosity. The angular separation between the components is $\rho = 34.3\arcsec$.

\textit{HD 96167:} This system was observed with Chandra's ACIS-I instrument but was not the main target of the observation. It is positioned $\approx 8.5 \arcmin$ from the optical axis. Since the angular separation between the components is $\approx 5.8 \arcsec$, we set the source extraction region to encompasses both components and collect most of the photon events that come from the system. The primary component is the planet host and also an evolved stellar object \citep{Fischer2005, Peek2009}. Therefore, we disregard this system from further analysis and give only the X-ray luminosity estimate for the unresolved system.

\textit{HD 101930:} The primary component was inside the FoV of all three EPIC detectors, whereas the secondary was only in the FoV of PN (in MOS1 and MOS2, the secondary was on the chip edge). Neither component was detected, therefore we estimated an upper limit to their X-ray luminosity assuming a coronal temperature of $\log_{10}T=6.477$ for both stars. The angular separation between the components is $\rho = 73.0\arcsec$.

\textit{HD 107148:} The system was observed with XMM-Newton (all three EPIC detectors) and Chandra X-ray observatory. The observation with XMM-Newton has a high background noise after approx. 26 ks of exposure time (the exposure time was 41 ks). We, therefore, used the shorter good time interval, where the detector experienced a low background signal. The primary component was detected, but with no hardness ratio estimate. The secondary component was not detected. We assumed for both components a coronal temperature of $\log_{10}T=6.477$ to estimate their X-ray luminosity. In the Chandra observation, the primary was detected, while the secondary was not. The secondary component is a white dwarf \citep{Mugrauer2016}. The angular separation between the components is $\rho = 35.0 \arcsec$.

\textit{HD 109749}: This system was detected by the Chandra ACIS-S instrument. The angular separation of the components is $\rho \approx 8.4 \arcsec$.

\textit{HD 178911:} This system was observed by Chandras ACIS-I instrument. It is a hierarchical triple star system. The resolved component is the planet host. Since both the unresolved binary system and the planet host were detected, we estimated the X-ray luminosity of the two components in the unresolved system. The binary is a G1-K1 pair \citep{Tokovinin2000}, therefore, we used the ratio of the expected X-ray luminosity of stellar corona for these spectral types and applied it to the luminosity we measured for the binary system. With this approach, we were able to use this system for further analysis. The angular separation between the planet host and the unresolved binary is $\rho \approx 16 \arcsec$. For Spearman's rank correlation calculation, we only used the percentile difference and tidal interaction strength parameters calculated for the AB pair.

\textit{HD 185269:} This system was observed with Chandra's ACIS-S instrument. The primary component, and the planet host of this system, is an evolved stellar object \citep{Johnson2006}, while the secondary component is an unresolved binary system Bab \citep{Ginski2016}. Although both components are marked as detected, this system was not used for further analysis. The angular separation between the projections of the components is $\rho \approx 4.5 \arcmin$.

\textit{HD 188015:} This system was observed with the ACIS-S instrument onboard the Chandra space observatory. Both components were detected. The angular separation between the components of this system is $\rho \approx 13 \arcsec$.

\textit{HD 189733:} This system was observed with Chandra's ACIS-S instrument. Both components were detected. The angular separation between the components is $\rho \approx 11.5 \arcsec$.

\textit{HD 190360:} The system was observed in two XMM-Newton observations. The primary is a weak X-ray source with no detection in the soft and hard bands. Therefore, we combined the two observations of the primary for a better SNR and set the coronal temperature to $\log_{10}T=6.477$ to estimate its X-ray luminosity. The secondary displayed a prominent stellar flare in observation no.\ 0304201101(obsID), and we, therefore, calculated its X-ray luminosity from the signal collected in the quiescent observation 0304202601(obsID).

\textit{HD 197037:} This system was observed by Chandra, with the ACIS-I instrument. The system projection on the instrument was at an off-axis angle of $\approx 9 \arcmin$. The angular separation between the components is $\approx 3.5 \arcsec$. Therefore, we chose a larger source extraction region that encompasses both components ($r_{extraction} \approx 10 \arcsec$), which gave us an estimate of the X-ray luminosity of the unresolved system.

\textit{HIP 116454:} This system was observed by Chandra but was not the main target of the observation. The projected position of this system on the sky had an off-axis angle of $\approx 18 \arcmin$ putting it on the edge of Chandra's FoV. Since the spatial resolution deteriorates significantly with increasing off-axis angle, the image of the system appears extended and unresolved. Therefore, we chose an extraction region that encompasses both sources. The extraction region radius is $\approx 22 \arcsec$, while the projected angular separation between the components is $\rho = 8.4 \arcsec$. Still, the system was marked as undetected by the ACIS-S camera. Apart from being unresolved, the stellar companion of the planet-hosting star has evolved from the main sequence \citep{Vanderburg2015,Mugrauer2019}, therefore, we disregarded this system from further analysis.

\textit{Kepler-444:} This system was observed with the Chandra space observatory. It is a hierarchical triple star system \citep{Campante2015}, where the secondary component is a spatially unresolved pair of M dwarfs. The projected separation of the A and the BC components is $\rho = 1.8 \arcsec$ ($\approx 70 AU$), making this system the most tightly bound in our sample. Neither of the components was detected, therefore we calculated the upper limits to their X-ray luminosities. 

\textit{Kepler-1008:} The two components of this system were observed with the MOS2 and PN detectors (their coordinates on MOS1 were outside the FoV) but were not detected. We estimated an upper limit to their X-ray luminosity. They are at an angular separation of $\rho = 13.4\arcsec$ from each other.

\textit{$\upsilon$ And (HD 9826):} This system was observed with XMM-Newton (all three EPIC detectors) and Chandra X-ray observatory. The XMM-Newton observation shows the primary as a bright X-ray source. For the second component, we assumed a coronal temperature of $\log_{10}T=6.477$ as the estimation of the radiation hardness ratio was not possible. In the Chandra observations, only the primary was in the FoV and was detected. The angular separation between the components is $\rho = 55.6\arcsec$.

\textit{WASP-8:} This system was observed by Chandra with ACIS-S. The primary was detected, while the secondary component was not. The projected angular separation between the components is $\rho = 4.5 \arcsec$.

\textit{WASP-18}: Both components were in the FoV of all three EPIC detectors but were not detected. Therefore, we estimated an upper limit to their X-ray luminosity. This system was also observed with Chandra: the primary component was again not detected, while the secondary component was projected between the ACIS-I chip array and was discarded from our analysis. The angular separation between the components is $\rho = 26.7 \arcsec$.

\textit{WASP-33:} This system was observed with the XMM-Newton telescope. It is a possible hierarchical triple star system \citep{Mugrauer2019}. The primary component is the planet host and has a close unresolved companion at $\rho=2\arcsec$. The wider companion is at an angular separation of $\rho=49.0\arcsec$. The wide companion was in the FoV of the PN detector, while the unresolved binary was in the FoV of MOS1 and PN. The secondary component was detected and we calculated its X-ray luminosity, while the unresolved binary was undetected and only an upper limit to its X-ray luminosity was estimated.

\textit{WASP-77:} This system was observed with the Chandra space observatory and the ACIS-S instrument. The two components of this system are at the angular separation of $\rho = 3.3 \arcsec$. The primary component was detected, while for the secondary component we estimated an upper limit for the X-ray flux and luminosity, assuming a coronal temperature of $\log_{10} T = 6.477$.

\textit{XO-2:} The system has two observations with XMM-Newton. Both components were detected in both observations (obsID: 0728970101 and 0728970201), but a radiation hardness ratio could not be estimated in each separate case. We, therefore, combined the appropriate observations to achieve a better SNR, but we were not able the calculate a hardness ratio. The coronal temperature of $\log_{10}T = 6.477$ was assumed for both components to calculate their X-ray luminosity. The angular separation between the components is $\rho = 31.0\arcsec$. Here, both stars are of the same spectral type and have planets: XO-2 N hosts a planet with $M_{pl} \approx 0.6 M_{Jup}$ at $a \approx 0.04 AU$, whereas XO-2 S has a 0.26$M_{Jup}$ planet at a distance of a = 0.13 AU. The XO-2 S star hosts a less massive, wider-orbiting planet and it can be expected that it does not exhibit a strong tidal pull onto its host star when compared to the XO-2 N system.

\section{Photon count conversion factors and Gaia parameters}

In table \ref{conversionfactorsXMM} are given the photon count conversion factors for each observation and camera taken with the XMM-Newton space observatory. In table \ref{conversionfactorsChandra} are given the conversion factors for the given observation cycle and CCD chip of the Chandra observatory. The photon count conversion factor was estimated with the WebPimms online tool and by calculating the radiation hardness ratio HR. If the source was not detected in the soft or hard passband, a coronal temperature of $\log_{10}T=6.477$ was assumed (see chapter \ref{conversionfactor} for more details). Table \ref{tab:gaia_params} shows the Gaia parameters used for stellar mass estimation of stars in our final sample.
\begin{table*}
\begin{tabular}{lllllrrrr}
\hline
\hline
system & obs ID & component &     detection &  HR &      {\it c}(MOS) &      {\it c}(PN) \\
\hline

    \multirow{2}{*}{16 Cyg} & \multirow{2}{*}{0823050101} &  AC &         BRIGHT &     -0.466  &  9.326602e-12 &  1.045563e-12 \\
 &  &  B* &     BRIGHT &     -0.460 &   9.284336e-12 &  1.044785e-12 \\
    \multirow{2}{*}{30 Ari} & \multirow{2}{*}{0075940101} & A &     BRIGHT &     -0.068  &  6.159409e-12 &  1.073431e-12 \\
     &  &  B*C &      BRIGHT &      0.111 &   6.047903e-12 &  1.096918e-12 \\
    \multirow{2}{*}{55 Cnc} & \multirow{2}{*}{0551020801} & A* &    FAINT &     -0.307 &    9.084168e-12 &  1.634601e-12 \\
     &  & B &  FAINT/NO HR &  /  &    9.084076e-12 &  1.634588e-12 \\
    \multirow{2}{*}{83 Leo} & \multirow{2}{*}{0551021201}  &  A &   FAINT &     -0.497 &    1.035328e-11 &  1.803968e-12 \\
     &   & B* &   BRIGHT &     -0.056 &    8.384539e-12 &  1.585091e-12 \\
    AS 205 &  0602730101  & A*B &     FAINT/NO HR &  / &    6.510630e-12 &  1.041020e-12 \\
    \multirow{2}{*}{GJ-15} &  \multirow{2}{*}{0801400301}  & A* &   BRIGHT &     -0.468 &  6.622050e-12 &  1.045839e-12 \\
     &    & B &         DETECTED &     -0.608 &    7.080730e-12 &  1.064492e-12 \\
  \multirow{4}{*}{HAT-P-16} & \multirow{2}{*}{0800733101} & A*B & NOT DETECTED &   /&   5.942612e-12 &  9.378096e-13 \\
   &  & C &  NOT DETECTED &       / &   5.942612e-12 &  9.378096e-13 \\
   & \multirow{2}{*}{0800730701} & A*B &     FAINT/NO HR &      / &   5.942612e-12 &  9.378096e-13 \\
   &  & C &    FAINT/NO HR &      / &  5.942612e-12 &  9.378096e-13 \\
  \multirow{4}{*}{HD 27442} & \multirow{2}{*}{0780510501} &  A* &     FAINT &     -0.676 &    7.328724e-12 &  1.073849e-12 \\
   &  & B &   FAINT/NO HR &      / &  6.510630e-12 &  1.041020e-12 \\
   & \multirow{2}{*}{0551021401} & A* &   FAINT/NO HR &   -0.676   &    1.193506e-11 & 2.000616e-12\\
   &  & B &  NOT DETECTED &    / &   9.084076e-12 &  1.634588e-12 \\
  HD 46375 & 0304202501 & A*B &    FAINT &     -0.561 &    6.920167e-12 &  1.058170e-12 \\
  \multirow{4}{*}{HD 75289} & \multirow{2}{*}{0304200501} & A* &   FAINT/NO HR &   / &   6.510630e-12 &  1.041020e-12 \\
   &  & B &    FAINT &     -0.300 &    6.310412e-12 &  1.044406e-12 \\
   & \multirow{2}{*}{0722030301} & A* &  NOT DETECTED &     / &   6.510630e-12 &  1.041020e-12 \\
   &  & B &      FAINT &     -0.300 &    6.310441e-12 &  1.044401e-12 \\
  \multirow{2}{*}{HD 101930} & \multirow{2}{*}{0555690301} & A* &  NOT DETECTED &     / & 5.942612e-12 &  9.378096e-13 \\
  &  & B &  NOT DETECTED &   / &  / &  9.378096e-13 \\
  \multirow{2}{*}{HD 107148} &  \multirow{2}{*}{0693010401} & A* &  FAINT/NO HR &  / &  5.942612e-12 &  9.378096e-13 \\
    & & B & FAINT/NO HR &    / &   5.942612e-12 &  9.378096e-13 \\
 \multirow{3}{*}{HD 190360} & 0304201101 & A* &  NOT DETECTED &    / &   6.510630e-12 &  1.041020e-12 \\
  & \multirow{2}{*}{0304202601} & A* &   FAINT/NO HR &    / &    6.510630e-12 &  1.041020e-12 \\
  &  & B &    FAINT &     -0.449 &    / &  1.043299e-12 \\
 \multirow{2}{*}{Kepler-1008} & \multirow{2}{*}{0550451901}  & A* &  NOT DETECTED &   / &    6.510630e-12 &  1.041020e-12 \\
 & & B & NOT DETECTED &   / &    6.510630e-12 &  1.041020e-12 \\
 \multirow{2}{*}{$\upsilon$ And} &  \multirow{2}{*}{0722030101} & A* &  BRIGHT &     -0.359 &  9.400024e-12 &  1.677780e-12 \\
 & & B &  FAINT/NO HR &  / &    9.084076e-12 &  1.634588e-12 \\
 \multirow{2}{*}{WASP-18} &  \multirow{2}{*}{0673740101} & A* &  NOT DETECTED & /  &   6.510630e-12 &  1.041020e-12 \\
 & & B &  NOT DETECTED &   / &   6.510630e-12 &  1.041020e-12 \\
 \multirow{2}{*}{WASP-33} & \multirow{2}{*}{0785120201} & A*B &  NOT DETECTED & / &  5.942612e-12 &   9.378096e-13 \\
 &  & C &   FAINT &      0.116 &  / &   1.042992e-12 \\
 \multirow{4}{*}{XO-2} & \multirow{2}{*}{0728970101} & S* &  FAINT/NO HR &  / &  6.510630e-12 &  1.041020e-12 \\
 & & N* &  FAINT/NO HR &  / &    6.510630e-12 &  1.041020e-12 \\
  & \multirow{2}{*}{0728970201} & S* &  FAINT/NO HR &  / &   6.510630e-12 &  1.041020e-12 \\
  &  & N* &  FAINT/NO HR &  / &    6.510630e-12 &  1.041020e-12 \\
\hline
\end{tabular}
\caption{The photon count conversion factors given in this table are calculated using the online tool WebPimms for observations made with the XMM-Newton space telescope. The conversion factor translating the photon counts detected with the MOS cameras into the X-ray flux has the same value for both MOS1 and MOS2.}
\label{conversionfactorsXMM}
\end{table*}

\begin{table*}
\begin{tabular}{llllrl}
\hline
\hline
     system &    component &  obsID &                 detection &  HR &                    {\it c} \\
\hline
     \multirow{6}{*}{16 Cyg} & \multirow{3}{*}{16647} &   B*   &  NOT DETECTED &        / &   3.141789e-11 \\
     &  &  A  &        FAINT &      0.863 &         1.422014e-11 \\
     & & C &      FAINT/NO HR &        / &   3.141789e-11 \\
      & \multirow{3}{*}{18756}&  B*   &  NOT DETECTED &        / &   3.141789e-11 \\
     &  &  A  &                  FAINT &      0.863 &         1.422014e-11 \\
      & & C  &           FAINT/NO HR &        / &   3.141789e-11 \\
     \multirow{2}{*}{55 Cnc} &  14401&   A*  &                  FAINT &     -0.865 &        7.527662e-12 \\
      &  14402 &  A* &    FAINT &     -0.865 &        7.527662e-12 \\
       AS 205 &  16327&   A*B  &                  FAINT &      0.918 &         4.080275e-12\\
       CoRoT-2&  10989 &    B  &  NOT DETECTED &        / &   1.141643e-11 \\
   \multirow{2}{*}{HAT-P-20}&   \multirow{2}{*}{15711}  &   A* &                  FAINT &      0.618 &        5.385769e-12 \\
   &  &  B  &           FAINT/NO HR &        / &   7.340064e-12 \\
   \multirow{2}{*}{HAT-P-22}&  \multirow{2}{*}{15105} &   A*  &  NOT DETECTED &        / &  2.096742e-11 \\
    &  &    B  &  NOT DETECTED &        / &  2.096742e-11 \\
    \multirow{2}{*}{HATS-65}&   3282 &   A*B  &  NOT DETECTED &        / &   9.127872e-12 \\
    &   9382 &   A*B  &  NOT DETECTED &        / &   1.787365e-11 \\
 HIP116454 &  19517 &   A*B  &  NOT DETECTED &        / &   5.099088e-11 \\
   \multirow{2}{*}{HD 46375}&  \multirow{2}{*}{15719} &   A*  &                  FAINT &     -0.177 &        8.042113e-12 \\
   &   &    B  &     FAINT &     -0.336 &        8.633701e-12 \\
   HD 96167 &   5817  &   A*B  &           FAINT/NO HR &        / &   1.563648e-11 \\
  \multirow{2}{*}{HD 107148} &  \multirow{2}{*}{13665} &   A*  &                  FAINT &     -0.213 &        5.851814e-12 \\
  &   &    B  &  NOT DETECTED &        / &  5.552321e-12 \\
  \multirow{2}{*}{HD 109749}&  \multirow{2}{*}{15720} &   A*  &    FAINT &     -0.333 &        8.621279e-12 \\
   &   & B &     FAINT &      0.252 &        6.533847e-12 \\
 \multirow{2}{*}{HD 178911}&  \multirow{2}{*}{13659} &   B*  &                  FAINT &     -0.155 &        5.720604e-12 \\
 &   & AC  &                  FAINT &      0.171 &        5.066729e-12 \\
  \multirow{2}{*}{HD 185269}&  \multirow{2}{*}{15721} &   A*  &                  FAINT &     -0.277 &        8.414531e-12 \\
  &   & Bab  &       FAINT &      0.126 &        6.927864e-12 \\
  \multirow{2}{*}{HD 188015}&  13667 &  A*  &                  FAINT &     -0.175 &        5.765516e-12 \\
   &  &  B  &           FAINT/NO HR &        / &  5.552321e-12 \\
 \multirow{12}{*}{HD 189733} &  \multirow{2}{*}{12340}&   A  &           BRIGHT &      0.258 &   4.854420e-12 \\
  &   &    B  &           BRIGHT &      0.187 &  4.978932e-12 \\
   &  \multirow{2}{*}{12341}&   A  &           BRIGHT &      0.258 &   4.854420e-12 \\
   &  &    B  &           BRIGHT &      0.187 &  4.978932e-12 \\
   &  \multirow{2}{*}{12342}&   A  &           BRIGHT &      0.258 &   4.854420e-12 \\
   &  &    B  &           BRIGHT &      0.187 &  4.978932e-12 \\
   &  \multirow{2}{*}{12343}&   A  &           BRIGHT &      0.258 &   4.854420-12 \\
   &  &    B  &           BRIGHT &      0.187 &  4.978932e-12 \\
   &  \multirow{2}{*}{12344}&   A  &           BRIGHT &      0.258 &   4.854420e-12 \\
   &  &    B  &           BRIGHT &      0.187 &  4.978932e-12 \\
   &  \multirow{2}{*}{12345}&   A  &           BRIGHT &      0.258 &   4.854420e-12 \\
  &   &    B  &           BRIGHT &      0.187 &  4.978932e-12 \\
 \multirow{4}{*}{HD 197037} &   7444&   A*B  &                  FAINT &      0.378 &        1.401012e-11 \\
   &   8598&   A*B  &      FAINT &      0.378 &        1.401012e-11 \\
   &   9770&   A*B  &      FAINT &      0.378 &        1.401012e-11 \\
   &   9771&   A*B  &      FAINT &      0.378 &        1.401012e-11 \\
 \multirow{2}{*}{Kepler-444} &  17733&   A*BC  &           FAINT/NO HR &        / &  1.203499e-11 \\
  &  20066 &   A*BC &           FAINT/NO HR &        / &  1.203499e-11 \\
    \multirow{4}{*}{$\upsilon$ And} &  10976 &   A*  &           BRIGHT &     -0.171 &   4.704921e-12 \\
    &  10977 &    A*  &           BRIGHT &     -0.171 &   4.704921e-12 \\
    &  10978 &    A*  &           BRIGHT &     -0.171 &   4.704921e-12 \\
    &  10979 &    A*  &           BRIGHT &     -0.171 &   4.704921e-12 \\
     \multirow{2}{*}{WASP-8}&  \multirow{2}{*}{15712} &   A*  &                  FAINT &      0.210 &        5.046927e-12 \\
      &  &  B  &  NOT DETECTED &        / &   5.608443e-12 \\
    WASP-18 &  14566&   A*  &  NOT DETECTED &        / &  2.096742e-11 \\
  \multirow{2}{*}{WASP-77} &  \multirow{2}{*}{15709} &   A* &                  FAINT &      0.363 &        6.185815e-12 \\
    & &   B  &  NOT DETECTED &        / &   7.340064e-12 \\
\hline
\end{tabular}

\caption{Given are the photon count conversion factors that are used to calculate the X-ray flux of stars observed with the Chandra space observatory.}
\label{conversionfactorsChandra}
\end{table*}

\begin{table*}
\begin{tabular}{cccccccc}
\hline
\hline
system & component & G & G corrected & $G-R_p$ & r[pc] & $M_G$ & RUWE \\
\hline
\multirow{3}{*}{16 Cyg} & B* & 6.0568 & 6.0566 & $0.4709 \pm 0.0019$ & $21.139 \pm 0.015$ & $4.4311 \pm 0.0012$ & 0.9414\\
& A & / & / & / & / & / & / \\
& C & / & / & / & / & / & / \\[0.2cm]
\multirow{2}{*}{55 Cnc} & A* & 5.7144 & 5.7297 & $0.6331 \pm 0.0028$ & $12.586 \pm 0.012$ & $5.5841 \pm 0.0250$ & 0.9865\\
& B & 11.6798 & 11.6617 & $1.2495 \pm 0.0016$ & $12.477 \pm 0.017$ & $11.1817 \pm 0.0032$ & 1.3065\\[0.2cm]
\multirow{2}{*}{CoRoT-2} & A* & 12.2489 & 12.2289 & $0.55659 \pm 0.0009$ & $213.283 \pm 2.457 $ & $5.2304 \pm 0.0022$ & 0.8829\\
& B & 15.4750 & 15.4447 & $1.1001 \pm 0.0056$ & $202.929 \pm 2.623$ & $8.9080 \pm 0.0281$ & 1.2250\\[0.2cm]
\multirow{2}{*}{GJ-15} & A* & 7.2162 & 7.2123 & $1.0340 \pm 0.0013$ & $3.562 \pm 0.001$ & $9.4537 \pm 0.0006$ & 0.9172\\
& B & 9.6774 & 9.6656 & $1.2051 \pm 0.0011$ & $3.561 \pm 0.001$ & $11.9077 \pm 0.0008$ & 1.0262\\[0.2cm]
\multirow{2}{*}{HAT-P-20} & A* & 10.9903 & 10.9743 & $0.7556 \pm 0.0018$ & $71.037 \pm 0.199$ & $6.7169 \pm 0.0061$ & 1.0556\\
& B & / & / & / & / & / & /\\[0.2cm]
\multirow{2}{*}{HD 46375} & A* & 7.6953 & 7.6899 & $0.5607 \pm 0.0020$ & $29.553 \pm 0.038$ & $5.3369 \pm 0.0028$ & 0.8831\\
& B & 11.2088 & 11.1921 & $1.0539 \pm 0.0029$ & $29.680 \pm 0.059$ & $8.8298 \pm 0.0045$ & 1.1117\\[0.2cm]
\multirow{2}{*}{HD 75289} & A* & 6.2052 & 6.2046 & $0.4120 \pm 0.0024$ & $29.116 \pm 0.024$ & $3.8839 \pm 0.0018$ & 0.8880\\
& B & / & / & / & / & / & /\\[0.2cm]
\multirow{2}{*}{HD 109749} & A* & 8.0245 & 8.0181 & $0.4655 \pm 0.0017$ & $63.082 \pm 0.295$ & $4.0185 \pm 0.0102$ & 1.0052\\
& B & / & / & / & / & / & /\\[0.2cm]
\multirow{3}{*}{HD 178911}& B* & 7.8670 & 7.8611 & $0.5072 \pm 0.0012$ & $40.973 \pm 0.046$ & $4.7986 \pm 0.0025$ & 0.9443\\
& A & / & / & / & / & / & /\\
& C & / & / & / & / & / & /\\[0.2cm]
\multirow{2}{*}{HD 188015} & A* & 8.0659 & 8.0593 & $0.4897 \pm 0.0013$ & $50.671 \pm 0.109$ & $4.5355 \pm 0.0047$ & 1.0189\\
& B & 15.4531 & 15.4229 & $1.2922 \pm 0.0040$ & $50.268 \pm 0.167$ & $11.9164 \pm 0.0074$ & 1.0643\\[0.2cm]
\multirow{2}{*}{HD 189733} & A* & 7.4143 & 7.4098 & $0.6063 \pm 0.0020$ & $19.764 \pm 0.013$ & $5.9304 \pm 0.0014$ & 0.9279\\
& B & 13.2055 & 13.1825 & $1.2572 \pm 0.0024$ & $19.711 \pm 0.019$ & $11.7089 \pm 0.0022$ & 1.1842\\[0.2cm]
\multirow{2}{*}{HD 190360} & A* & 5.5336 & 5.5543 & $0.5000 \pm 0.0029$ & $16.007 \pm 0.016$ & $4.5328 \pm 0.0024$ & 0.8292\\
& B & 12.7967 & 12.7750 & $1.3046 \pm 0.0028$ & $15.97 \pm 0.015$ & $11.7584 \pm 0.0021$ & 1.1162\\[0.2cm]
\multirow{2}{*}{$\upsilon$ And} & A* & 3.8985 & 3.9869 & $0.3503 \pm 0.0081$ & $13.405 \pm 0.063$ & $3.3505 \pm 0.0121$ & 0.8432\\
& B & 12.5126 & 12.4918 & $1.2710 \pm 0.0021$ & $13.471 \pm 0.016$ & $11.8449 \pm 0.0027$ & 1.1438\\[0.2cm]
\multirow{2}{*}{WASP-8} & A* & 9.6125 & 9.6009 & $0.5074 \pm 0.0020$ & $89.961 \pm 0.363$ & $4.8307 \pm 0.0088$ & 1.0193\\
& B & 13.6980 & 13.6733 & $1.1191 \pm 0.0081$ & $90.468 \pm 0.364$ & $8.8909 \pm 0.0088$ & 1.2506 \\[0.2cm]
\multirow{2}{*}{WASP-77} & A* & 10.0966 & 10.0835 & $0.5110 \pm 0.0029$ & $105.166 \pm 1.196$ & $4.9741 \pm 0.0247$ & 1.1304 \\
& B & / & / & / & / & / & / \\[0.2cm]
\multirow{2}{*}{XO-2} & S* & 10.9278 & 10.9121 & $0.5567 \pm 0.0009$ & $151.398 \pm 0.95$ & $5.0115 \pm 0.0136$ & 0.9168\\
& N* & 10.9718 & 10.9559 & $0.5688 \pm 0.0013$ & $154.273 \pm 1.446$ & $5.0145 \pm 0.0203$ & 0.9437\\
\hline
\end{tabular}
\caption{Given are the (corrected) Gaia apparent magnitude, the $G-R_p$ color, the calculated Gaia absolute magnitude, and the renormalised unit weight error (RUWE) for each star of our final sample. The uncertainties given for the color and absolute magnitude were calculated via the error propagation function and are relevant for the estimation of the stellar evolutionary status: on the MS or evolved. The stars which do not have their magnitudes, color, and RUWE given here did not pass some of the quality assessment and their spectral type/mass was acquired from the literature as given in Tab. \ref{tab:gaiamasses}.}
\label{tab:gaia_params}
\end{table*}

\bsp	
\label{lastpage}
\end{document}